\documentclass[%
 notitlepage,
 twocolumn,
superscriptaddress,
 amsmath,amssymb,
 aps,
 pra,
]{revtex4-1}
\usepackage{amsfonts}
\usepackage{graphicx}
\usepackage{physics}
\usepackage{dcolumn}
\usepackage{bm}
\usepackage{bbold}
\usepackage{soul}
\usepackage{cancel}
\usepackage{latexsym}
\usepackage{color}
\usepackage{xcolor}
\usepackage{dsfont}
\usepackage{enumerate}
\usepackage{comment}
\usepackage{ulem}
\usepackage{hyperref}
\hypersetup{
citecolor=blue,
colorlinks=true,
urlcolor=magenta,
}

\newcommand{\ICFO}{ICFO - Institut de Ciencies Fotoniques, The Barcelona Institute of Science and Technology, 08860 Castelldefels, Barcelona, Spain}
\newcommand{\ICREA}{ICREA, Pg. Lluís Companys 23, 08010 Barcelona, Spain}

\newcommand{\IOPPAS}{Institute of Physics PAS, Aleja Lotnikow 32/46, 02-668 Warszawa, Poland}

\begin{document}

\title{
Accelerating many-body entanglement generation by dipolar interactions in the Bose-Hubbard model
}

\author{Marlena Dziurawiec}
\affiliation{\IOPPAS}

\author{Tanaus\'u Hern\'andez Yanes}
\affiliation{\IOPPAS}

\author{Marcin P\l{}odzie\'n}
\affiliation{\ICFO}

\author{Mariusz Gajda}
\affiliation{\IOPPAS}

\author{Maciej Lewenstein}
\affiliation{\ICFO}
\affiliation{\ICREA}

\author{Emilia Witkowska}
\affiliation{\IOPPAS}

\date{\today}

\begin{abstract}
The spin squeezing protocols allow the dynamical generation of massively correlated quantum many-body states, which can be utilized in entanglement-enhanced metrology and technologies. We study a quantum simulator generating twisting dynamics realized in a two-component Bose-Hubbard model with dipolar interactions. We show that the interplay of contact and long-range dipolar interactions between atoms in the superfluid phase activates the anisotropic two-axis counter-twisting mechanism, accelerating the spin squeezing dynamics and allowing the Heisenberg-limited accuracy in spectroscopic measurements.
\end{abstract}

\maketitle

\section{Introduction}
The second Quantum Revolution's main objective lies in multipartite entangled states: their production, storage, certification, and application. Such states, i.e., many-body entangled and many-body Bell correlated states, are essential resources for quantum-based technologies and quantum-enhancement metrology \cite{Acin_2018,Eisert2020,Kinos2021,Laucht_2021,Zwiller2022,Fraxanet2022}. As such, a general protocol allowing the controlled generation of such states is an extensive research direction in modern quantum science.
Spin squeezing represents such a protocol paving the way for high-precision measurements, allowing overcoming the shot-noise limit \cite{Kitagawa1993,Wineland1994}, generate many-body entangled
\cite{Fadel2018,PhysRevLett.122.173601,Hosten2016, Pedrozo2020, Bao2020},  and many-body Bell correlated states ~\cite{Tura1256,schmied2016bell,Aloy2019, Baccari2019, Tura2019, PRXQuantum.2.030329, Plodzien2022}. 
The spin squeezing applies to a system of $N$  quantum in two internal states corresponding to a
spin-1/2 degree of freedom, and further described by the collective spin of the quantum number $S = N/2$. The uncertainty of spectroscopic measurements is  $\xi/\sqrt{N}$, where 
  \begin{equation}\label{eq:ssqparameter}
    \xi^2 = \frac{N\Delta^2\hat{S}_{\perp\min}}{\langle S\rangle^2},
\end{equation}
is the spin squeezing parameter, and $\Delta^2\hat{S}_{\perp\min}$ is the minimal variance in the plane perpendicular to total spin vector. The spin squeezing is a witness of entanglement-depth,  i.e. quantum state is not $k$-producible, when  $\xi<1/k$ \cite{Sorensen2001_Nature,Pezze2009,Hyllus2012,Toth2012}.

The paradigmatic theoretical models realizing spin squeezing through unitary evolution are given by the so-called One-Axis Twisting (OAT), and Two-Axis Counter-Twisting (TACT) Hamiltonians \cite{Kitagawa1993}. The OAT Hamiltonian has the form of the non-linear operator, often cast as
\begin{equation}
    \hat{H}_{\rm OAT} = \hbar\chi\hat{S}_z^2,
\end{equation}
where $z$ is the twisting axis, and $\chi^{-1}$ is the time-scale on which spin squeezing parameter has the lowest value $\xi^{2}_{\rm best}$. The lowest value of the squeezing parameter scales with particle numbers, and for OAT it is $\xi_{\rm best}\propto N^{-1/3}$ at $\chi t_{\rm best} \simeq N^{-2/3}$ \cite{Kitagawa1993}. 
The TACT Hamiltonian reads \begin{equation}
    \hat{H}_{\rm TACT} = \hbar\chi(\hat{S}^2_z - \hat{S}^2_x),
\end{equation}
where the clockwise and counter-clockwise twisting take places around two orthogonal axes $z$ and $x$. 
The advantage of the TACT is that it gives the Heisenberg limited level of the best squeezing, namely $\xi_{\rm best}\propto N^{-1/2}$.
In addition, the time scale of the best squeezing is accelerated with respect to OAT, and it is given by $\chi t_{\rm best} \sim N^{-1}\log(2 N)$~\cite{PhysRevA.92.013623} .

Realizing quantum simulators of OAT or TACT dynamics is essential for quantum enhancement metrology. Ultra-cold atoms form a perfect platform for quantum simulators mimicking such twisting dynamics. OAT has been realized with Bose-Einstein condensates utilizing atom-atom collisions \cite{Sorensen1999PRL,Sorensen2001,Treutlein2010,Oberthaler2010,Chapman2012,PhysRevLett.125.033401},  and atom-light interactions~\cite{PhysRevLett.104.073602,PhysRevLett.105.080403}. Another research directions are ultra-cold platforms simulating the Hubbard and Heisenberg models, which generate twisting dynamics. In the case of bosons, twisting dynamics is induced by atom-atom collisions~\cite{Kajtoch2018,PhysRevResearch.1.033075, Plodzien2020, PhysRevResearch.3.013178}, while for spinful fermions a synthetic spin-orbit coupling is necessary \cite{Wall2016,Ye2017,Kolkowitz2017,Bromley2018,He2019,Mamaev2021,Hernandez2022}. 
Finally, twisting dynamics can be activated with long-range interacting
bosons, what provides a platform for spin squeezing simulators by casting original Hamiltonian onto long-range interacting spin-chain
\cite{Civitarese2010,Perlin2020,Bilitewski2021,Roscilde2021,Wu2022,Comparin2022_a,Comparin2022_b}.

In this work, we propose a quantum simulator
for the TACT model realized in a one-dimensional two-component Bose-Hubbard model in the superfluid phase,
considering both contact and dipolar interactions. With the help of full many-body dynamics and an
effective two-mode model (TMM) description, we show that the realized
squeezing dynamics capture properties of the anisotropic
TACT model where the clockwise and counter-clockwise twisting take place with different rates. Next, we show that scaling with the system size of the best
squeezing parameter and best squeezing time is equivalent to the scaling obtained
for the TACT model.
Our results show the significant acceleration of the spin squeezing dynamics by dipolar interactions, which is an essential effect from the experimental point of view

The paper is organized as follows. In Sec.\ref{sec:System} we introduce the considered model. Starting with general many-body description of the system we provide an effective two mode model accounting for both contact and long-range dipolar interactions. Next, in Sec.\ref{sec:PhaseSpace} we perform analysis of the mean-field phase space of the anisotropic TACT model. In Sec.\ref{sec:Scaling} with the help of Bogoliubov-Born-Green-Kirkwood-Yvon (BBGKY) hierarchy of equations~\cite{PhysRevA.64.013605, PhysRevA.65.053819}, we analyze the scaling of the best squeezing and the best squeezing time with the system size. We conclude in Sec.\ref{sec:Conclusions}.

\section{Exact and effective models}
\label{sec:System}

We consider $N$ bosonic atoms in the two internal states $\ket{\uparrow}$, $\ket{\downarrow}$ which corresponds to the ensemble of $N$ spin-1/2 particles (qubits).
The atoms are described by the following Hamiltonian:
\begin{align}
    &\hat{H} = \hat{H}_0 + \hat{H}_{\rm d}, \label{eq:Hgeneral}\\
    &\hat{H}_0 =\int d^3 \boldsymbol{r}\hat{\Psi} ^\dagger(\boldsymbol{r}) \left(-\frac{\hbar^2 \nabla ^2}{2m} +V_{\rm latt}\right)\hat{\Psi}(\boldsymbol{r}),  \\
    &\hat{H}_{\rm d} = \int d^3 \boldsymbol{r}_1 \int d^3 \boldsymbol{r}_2 \hat{\Psi} ^\dagger(\boldsymbol{r}_1)\hat{\Psi} ^\dagger(\boldsymbol{r}_2) V_{12}\hat{\Psi} (\boldsymbol{r}_2)\hat{\Psi} (\boldsymbol{r}_1),
    \label{eq:Hdipgeneral}
\end{align}
where the vector of bosonic field operators is $\hat{\Psi}^T(\boldsymbol{r})=(\hat{\Psi}_{\uparrow}({\bf r}), \hat{\Psi}_{\downarrow}(\boldsymbol{r}))$ with $\hat{\Psi}_{\sigma}(\boldsymbol{r})$ describing an atom at the position $\boldsymbol{r}$ in the state $\sigma = \uparrow, \downarrow$. The interaction potential
$V_{12}$ is a sum of two terms,  $V_{12}= V_{\rm c}+ V_{\rm d}$, the short range contact interaction
\begin{align}
    V_{\rm c} &= \frac{4\pi\hbar^2a_s}{m}\delta\left(\boldsymbol{r}_1-\boldsymbol{r}_2\right),
\end{align}
and the long range dipolar interaction   
\begin{align}    \label{eq:dip_int_base}
    V_{\rm d} &= \frac{\boldsymbol{\mu}_1 \cdot \boldsymbol{\mu}_2}{|\boldsymbol{r}_1 - \boldsymbol{r}_2|^3} - \frac{3[\boldsymbol{\mu}_1 \cdot (\boldsymbol{r}_1 - \boldsymbol{r}_2) ][\boldsymbol{\mu}_2 \cdot (\boldsymbol{r}_1 - \boldsymbol{r}_2) ]}{|\boldsymbol{r}_1 - \boldsymbol{r}_2|^5},
\end{align}
where $\boldsymbol{\mu}_{1,2}$ is the dipole moment, $m$ is the atomic mass and $a_s$ is the s-wave scattering length.

The atoms are loaded into the one-dimensional optical lattice potential $V_{\rm latt}= V_0 \sin^2 {k x}$, where $k=2 \pi/\lambda_{\rm latt}$ is a wave-vector associated with the lattice wave-length $\lambda_{\rm latt}$. We consider the unit filling, so the number of lattice sites $M$ equals the total number of atoms $N$ ($M=N$). We assume the atoms are in the superfluid phase and occupy the lowest Bloch band. In the tight-binding approximation, the field operators is conveniently expanded in the basis of the Wannier functions, and the system Hamiltonian (\ref{eq:Hgeneral}) reduces to the two-component Bose-Hubbard model (BHM) extended by the dipolar term, namely
\begin{equation}\label{eqn:H_dBH}
    \hat{H} = \hat{H}_{\rm BH} + \hat{H}_{\rm d} \equiv \hat{H}_{\rm dBH}.
\end{equation}
The two-component Bose-Hubbard Hamiltonian $\hat{H}_{BH}$ reads
\begin{align}
\label{eq:full_ham}
 &\hat{H}_{\rm BH}=-J \sum_{\sigma = \uparrow, \downarrow}
\sum_{j}
\left(
\hat{a}^\dagger_{\sigma,j} \hat{a}_{\sigma,j+1} 
+ \hat{a}^\dagger_{\sigma,j} \hat{a}_{\sigma,j-1}
\right) \nonumber \\
&+ \sum_{j}\left(\frac{U_{\uparrow\uparrow}}{2}\hat{n}_{j\uparrow} (\hat{n}_{j\uparrow}-1)+ \frac{U_{\downarrow\downarrow}}{2}\hat{n}_{j\downarrow}(\hat{n}_{j\downarrow}-1)+ U_{\uparrow\downarrow}\hat{n}_{j\uparrow}  \hat{n}_{j\downarrow} \right), 
\end{align}
where $\hat{a}_{\sigma,j}$ and $\hat{n}_{\sigma,j}$  are the  on-site annihilation and number operators of atoms in the state $|\sigma\rangle$ at the site $j$.
The hopping amplitude $J$ does not depend on the spin state $\sigma$, while the interaction coefficients $U_{\sigma\sigma'}$ contain the contributions of both the on-site contact and the on-site dipolar interaction~\cite{nonstandard}.
The dipolar interaction term $\hat{H}_{d}$ reads
 \begin{align}
 \hat{H}_{\rm d} 
 &=
     \sum_{j,k\ne j}\frac{\gamma^2}{4|j - k|^3}(
     \hat{S}_{z,j}\hat{S}_{z,k}-2\hat{S}_{x,j}\hat{S}_{x,k} +\hat{S}_{y,j}\hat{S}_{y,k}),
 \label{eq:dipolgeneralM}
\end{align}
where the dipole moment was associated with the spin operators, $\boldsymbol{\mu} = - \gamma \boldsymbol{S}$, and where the on-site spin operators are
$\hat{S}^{+}_j = \hat{a}^\dagger_{\uparrow,j}\hat{a}_{\downarrow,j}$, $\hat{S}^{-}_j =\hat{a}^\dagger_{\downarrow,j}\hat{a}_{\uparrow,j}$, $\hat{S}^{\pm}_j = \hat{S}_{x,j} \pm i\hat{S}_{y,j}$ and $\hat{S}_{z,j} = (\hat{n}_{j,\uparrow}-\hat{n}_{j,\downarrow})/2$, while the collective spin operators read $\hat{S}_x  = \frac{1}{2}\sum_j \hat{S}_{x,j}$, $\hat{S}_y  = \frac{1}{2 i}\sum_j \hat{S}_{y,j}$, and $\hat{S}_z = \frac{1}{2}\sum_j \hat{S}_{z,j}$.
The range of dipole potential extends over several lattice sites under typical experimental conditions. Therefore, it is approximately constant on scales comparable to the spatial localization of Wannier functions. Under this condition the dipolar part of the Hamiltonian can be simplified in the form (\ref{eq:dipolgeneralM}), see Appendix~\ref{app:derivation-dipol} for derivation.
 
We consider the dynamical generation of spin squeezed states from an initial spin coherent state when the system is in the superfluid phase, $U_{\sigma\sigma} \ll J$, and contact interactions compete with the long-range one. Such the regime corresponds to the situation when the wave-functions of atoms are delocalized over the entire lattice and the condensate fraction, 
$f_c \equiv \frac{1}{N^2}\sum_{i,j} \sum_{\sigma=\uparrow,\downarrow} \langle \hat{a}^{\dagger}_{\sigma,j}\hat{a}_{\sigma,j} \rangle$,
approximately equals one. 

The two-component Bose-Hubbard model can simulate the OAT dynamics via contact interactions among bosons in the superfluid phase~\cite{Plodzien2020}. 
Here we show that the system can simulate the anisotropic TACT dynamics when dipolar interactions between the bosonic atoms are taken into account. 
To understand why the twisting mechanism is simulated by the system Hamiltonian (\ref{eqn:H_dBH}) we perform the following analysis. 
First, we consider $\hat{H}_{\rm d BH}$ in the quasi-momentum representation by using the Fourier transforms, $\hat{a}_{\sigma,j}=\frac{1}{\sqrt{N}}\sum_n e^{i \frac{2 \pi}{M} j n} \hat{a}_{\sigma,q_n}$ and $\hat{S}_{\sigma,j}=\frac{1}{\sqrt{N}}\sum_n e^{i \frac{2 \pi}{M} j n} \hat{S}_{\sigma,q_n}$, where the quasi-momentum reads $q_n = \frac{2 \pi}{M} n$ for $n=0,1,2,\cdots,N-1$. Next, by keeping the zero-momentum mode contributions only $q_n=0$ one can show that $\hat{H}_{\rm d BH}$ reduces to the effective model that is a sum of two terms
\begin{equation}\label{H_dbH0}
    \hat{H}_{{\rm dBH}, q_n=0} = \hat{H}_{{\rm BH},q_n=0} + \hat{H}_{{\rm d},q_n=0}.
\end{equation}

The first term, $\hat{H}_{\rm BH,n=0}$, comes from the zero quasi-momentum mode of the Bose-Hubbard Hamiltonian,
\begin{align}\label{H_OAT}
    \hat{H}_{{\rm BH},q_n=0} &= -2J\hat{N}_{q_n=0} + \Omega_{NN} \hat{N}^2_{q_n=0} \nonumber \\
    &+ \Omega_{SN} \hat{S}_{z,q_n=0}\hat{N}_0+ \Omega_{SS} \hat{S}_{z,q_n=0}^2,
\end{align}
where
\begin{align}
    \Omega_{NN} &= \frac{U_{\uparrow\uparrow}+U_{\downarrow\downarrow}+ 4U_{\uparrow\downarrow}}{8N},\\
    \Omega_{SN} &= \frac{U_{\uparrow\uparrow}-U_{\downarrow\downarrow}}{2N},\\
    \Omega_{SS} &= \frac{U_{\uparrow\uparrow}+U_{\downarrow\downarrow}-2U_{\uparrow\downarrow}}{2N},
\end{align}
and realizes OAT dynamics \cite{Plodzien2020}.
The second term, $\hat{H}_{d,q_n=0}$, consists of zero momentum component of the dipolar interaction:
\begin{align}\label{H_d}
   \hat{H}_{{\rm d},q_n=0} 
   & = 2\frac{\gamma^2h^{(3)}_{\left \lfloor{N/2}\right \rfloor }}{N} \left(
   \hat{S}^2_{z,q_n=0} - 3\hat{S}^2_{x,q_n=0}\right),
\end{align}
with $h^{(3)}_{\left \lfloor{N/2}\right \rfloor} = \sum_{d=1}^{{\left \lfloor{N/2}\right \rfloor}} 1/d^3$. 
Finally, by collecting the particular terms in \eqref{H_dbH0} we obtain
\begin{equation}
    \hat{H}_{{\rm dHB}, q_n=0 } = \frac{U-U_{\uparrow\downarrow}}{N} \hat{S}^2_{z, q_n=0} - 6\frac{\gamma^2h^{(3)}_{\left \lfloor{N/2}\right \rfloor }}{N} \hat{S}^2_{x, q_n=0},\label{eq:TACTquasi0}
\end{equation}
where we neglected the constant energy terms assuming $\hat{N}_0=N$ and $U_{\uparrow\uparrow}=U_{\downarrow\downarrow}=U$.
Note here, the zero quasi-momentum component of the spin operators correspond to the collective spin operators in the position representation, namely $\hat{S}_{\beta, q_n=0}=\frac{1}{\sqrt{N}}\sum_j  \hat{S}
_{\beta,j}$ with $\beta=x,y,z$. As such, we can replace
$\hat{S}_{\beta, 0}$ by the collective spin operators $\hat{S}_\beta$ in (\ref{eq:TACTquasi0}). Taking this into account, we identify the effective two-mode model (TMM):
\begin{equation}
\label{eqn:2m}
    \hat{H}_{\rm TMM}=\hbar \chi \left( \hat{S}_z^2 - \eta \hat{S}_x^2\right)
\end{equation}
which is the anisotropic TACT with
\begin{align}
    \hbar \chi &=\frac{U-U_{\uparrow\downarrow}}{N}\\
    \hbar \chi\eta&=6\frac{\gamma^2h^{(3)}_{\left \lfloor{N/2}\right \rfloor }}{N}
\end{align}
where $\eta$ is the anisotropy parameter and $\chi$ sets the energy scale. In the two limit cases $\eta=0$ and $\eta=1$ the effective model (\ref{eqn:2m}) reduces to the OAT and TACT model, respectively.

\begin{figure}
    \centering
    \begin{picture}(120, 360)
    \put(-44,-7){\includegraphics[width=.89\linewidth]{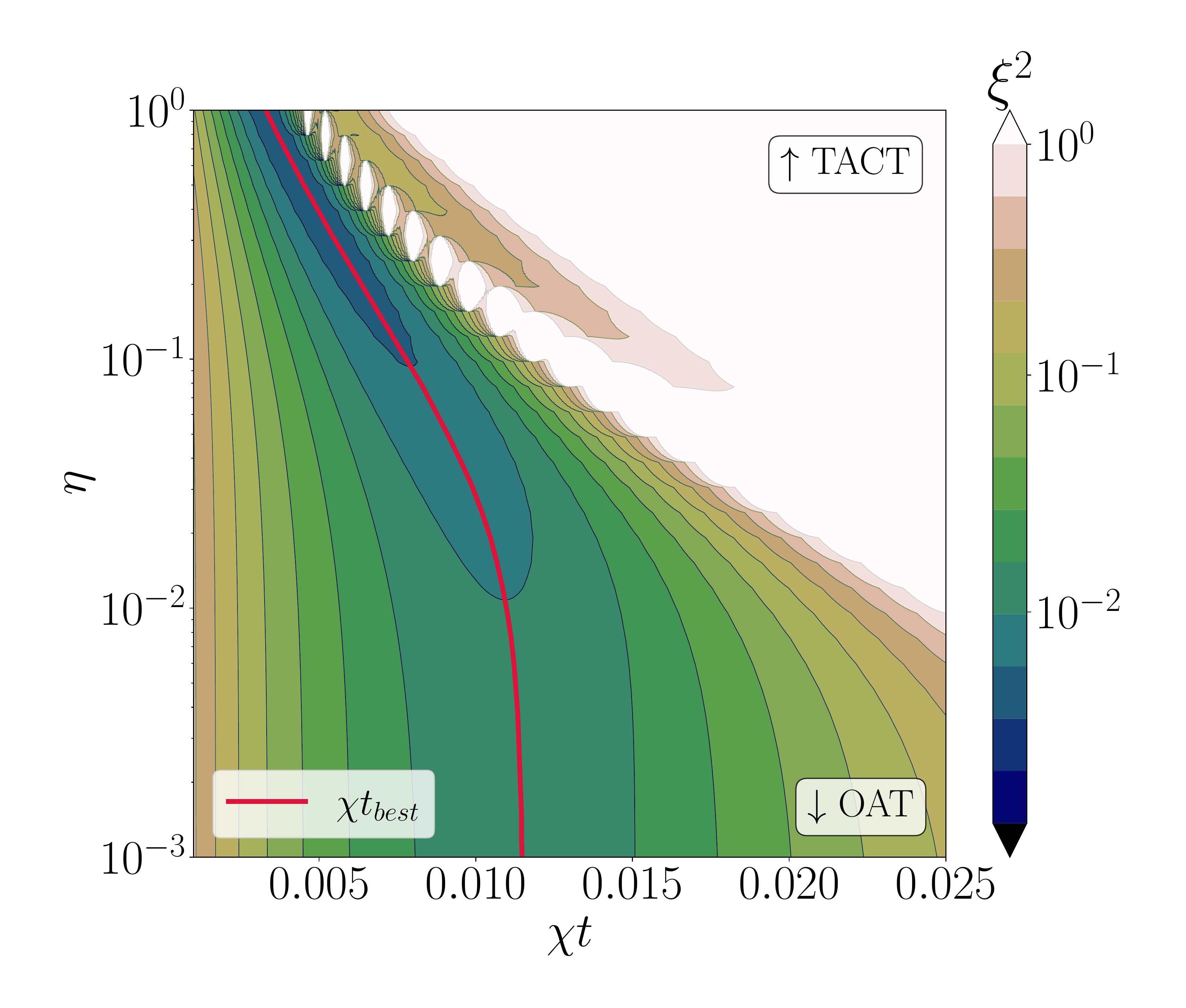}}
    \put(-40,170){\includegraphics[width=.80\linewidth]{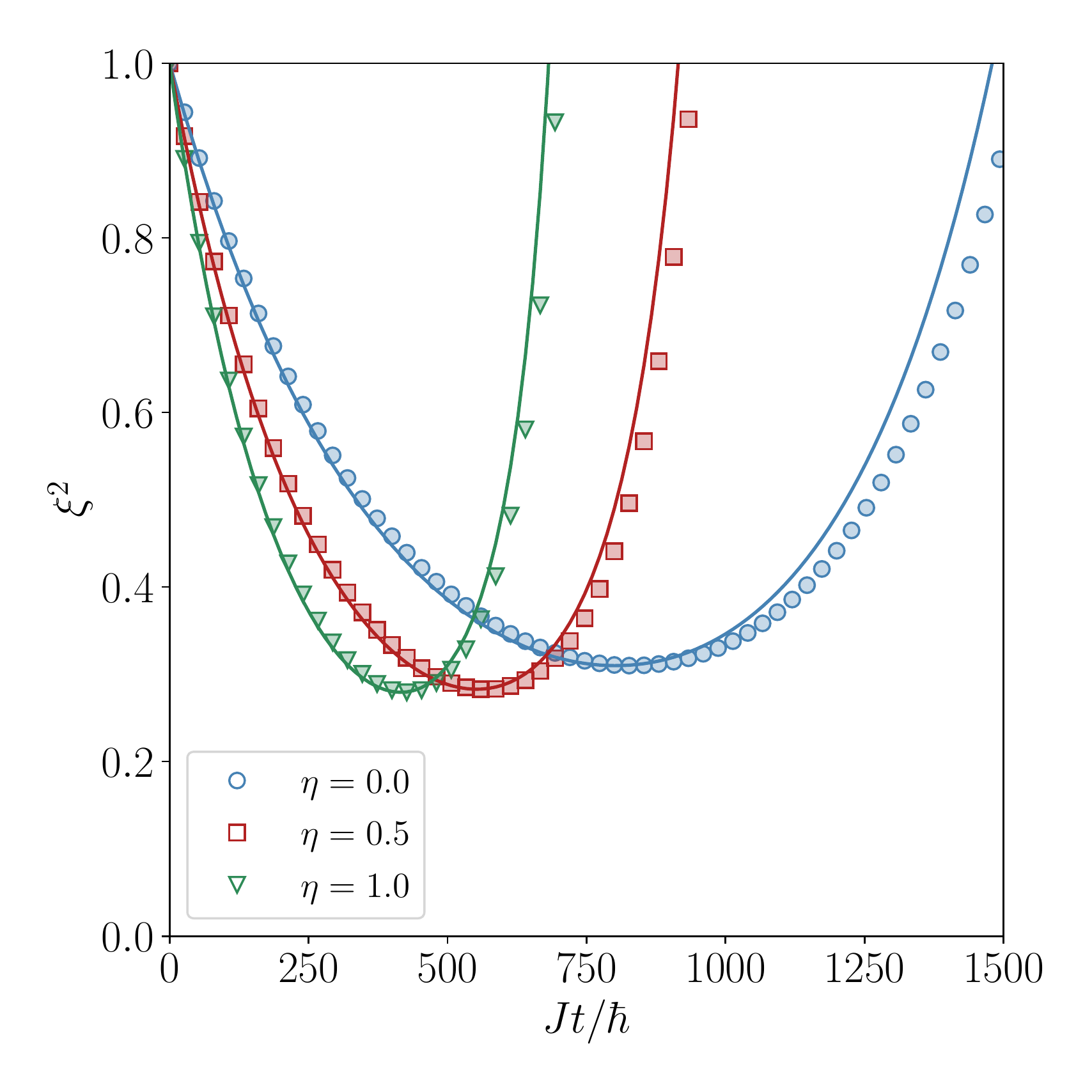}}
    \put(-40,351){(a)}
    \put(-40,155){(b)}
    \end{picture}
    \caption{(a) Time evolution of the spin squeezing parameter $\xi^2$ defined in (\ref{eq:ssqparameter}) for different values of anisotropy parameter~$\eta = \{0.0, 0.5, 1.0\}$ (lines from right to left). The two limiting cases, i.e. $\eta = 0$ and $\eta = 1$ correspond to OAT and TACT dynamics, respectively. 
    Points correspond to the results from the exact many-body numerical simulation of $\hat{H}_{\rm dBH}$ given by \eqref{eqn:H_dBH}, while  solid lines to the numerical results from the effective two-mode model \eqref{eqn:2m} when $N=M=10$, $U=0.01$, $J=1.0$ and $U_{\uparrow\downarrow}=0.95U$.
    (b) Color encoded values of the spin squeezing parameter $\xi^2$ versus $\chi t$ and $\eta$ obtained from the numerical simulations of the two mode model (\ref{eqn:2m}) for $N = 10^3$ atoms. The solid red line indicates the best squeezing time.}
    \label{fig:fig1}
\end{figure}

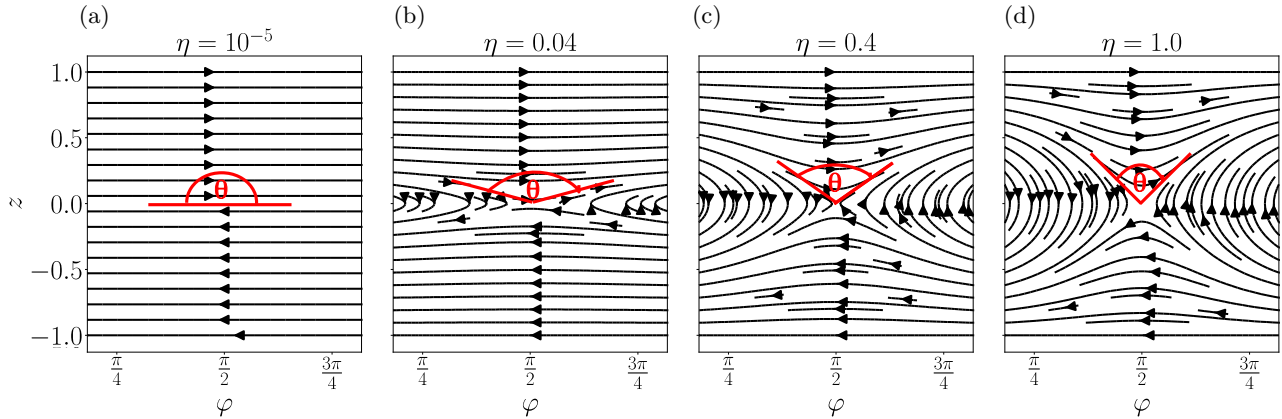
\begin{figure*}
	\centering
	\centering
    \begin{picture}(240, 180)
    \put(-140,-7){\includegraphics[width=\linewidth]{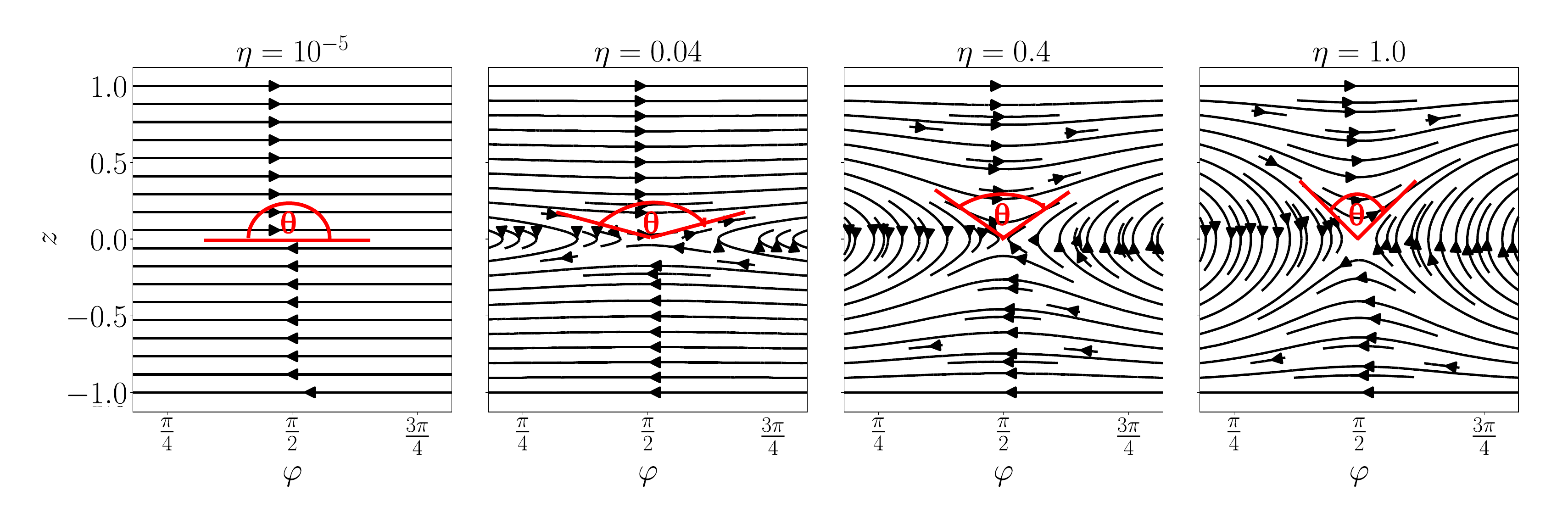}}
    \put(-100,152){(a)}
    \put(19,152){(b)}
    \put(132,152){(c)}
    \put(250,152){(d)}
    \end{picture}
	\caption{
		Mean-field phase portraits indicating geometrical representations of trajectories and directions of evolution for the effective two-mode model~(\ref{eqn:2m}) for the anisotropy parameter $\eta = \{10^{-5}, 0.04, 0.4, 1.0\}$ (from left to right).
		The angles $\theta$ between in-going and out-going flow along constant energy lines are marked by red (see main text). 
		The corresponding values of angles $\theta$ are approximately $\pi$, $0.87\pi$, $0.64\pi $, $0.5\pi$ (from left to right). 
	}
	\label{fig:fig2}
\end{figure*}

In Fig.\ref{fig:fig1} (a) we show spin squeezing parameter (\ref{eq:ssqparameter}) obtained from the exact many-body numerical simulation of the system dynamics under the dipolar Bose-Hubbard Hamiltonian \eqref{eqn:H_dBH} and the effective two-mode model \eqref{eqn:2m}, see Appendix~\ref{app:TMMnumerics} for more details concerning numerical simulations. For the chosen set of parameters, the condensate fraction is approximately one, $f_c \approx 1$, at the time scale corresponding to the best squeezing, which justifies our two-mode approximation. The overall agreement between the two models can be noticed. 
The acceleration of the squeezing dynamics is visible by increasing the value of the anisotropy parameter.
In Fig.\ref{fig:fig1}(b) we present a variation of the spin squeezing parameter $\xi^2$ in time and anisotropy parameter $\eta$ from the two-mode model (\ref{eqn:2m}) for $N=10^3$. One can observe the two limiting cases corresponding to OAT and TACT dynamics for $\eta = 0$ and $\eta = 1$, respectively. We observe a smooth transition between OAT and TACT dynamics in the intermediate region.

In the next paragraph, we provide an intuitive explanation for the OAT-TACT crossover with the help of phase portrait analysis of the two-mode model~\eqref{eqn:2m}. In Section \ref{sec:Scaling} we derive the scaling of the best squeezing and the best squeezing time with $N$ showing the acceleration of squeezing dynamics by the dipolar interactions.

\section{Mean-field phase portraits}\label{sec:PhaseSpace}

The activation of the TACT dynamics by dipolar interactions can be intuitively explained by analyzing the structure of the mean-field phase space of the two-mode model (\ref{eqn:2m}). 
It is a good navigator for the dynamical spin squeezing~\cite{PhysRevA.92.013623} as the eigenstates of quantum Hamiltonian localize on classical phase space energy contours~\cite{PhysRevA.79.013608} and quantum evolution distinguishes between stable and unstable classical fixed-points~\cite{PhysRevA.77.013614}.

The analysis of the mean-field phase space is performed by replacing the annihilation and creation operators by complex numbers~\cite{PhysRevLett.79.4950},
$\hat{a}  \rightarrow \sqrt{N\rho_a}e^{i\phi_a}$, 
$\hat{b}  \rightarrow \sqrt{N\rho_b}e^{i\phi_b}$ what transforms the spin operators to
$\hat{S}_x \rightarrow N\frac{\sqrt{1-z^2}}{2}\cos\phi$, 
$\hat{S}_y \rightarrow N\frac{\sqrt{1-z^2}}{2}\sin\phi$,
$\hat{S}_z \rightarrow \frac{N}{2}z$.
This allows introducing the new canonical variables $z=\rho_a-\rho_b$ and $\phi=\phi_a-\phi_b$. 
The Hamiltonian \eqref{eqn:2m} takes the form of the energy functional $\epsilon(\phi,z)$:
\begin{equation}\label{eq:epsilon}
\epsilon(\phi,z) = \frac{N}{4}z^2 - \frac{\eta N}{4}(1-z^2)\cos^2 \phi.
\end{equation}
Equations of motion for the canonical position $\phi$ and
the conjugate momentum $z$ are set by the Hamilton equations:
\begin{align}
\label{eq:mf}
    \dot{\phi}&=\frac{\partial\epsilon(\phi,z)}{\partial z} = \frac{N}{2}z +\frac{\eta N}{2}z\cos^2 \phi \nonumber\\
    \dot{z}&=-\frac{\partial\epsilon(\phi,z)}{\partial \phi} =  -\frac{\eta N}{2}(1-z^2)\cos \phi\sin \phi.
\end{align}

In the following we will analyze the topology of phase portraits which are a geometrical representation of trajectories of a dynamical system in the phase space.
In our case, trajectories are tangent to the velocity field $(\dot{\phi}, \dot{z})$.
The phase portrait consists of fixed points or closed orbits corresponding to a steady state, and satisfies $(\dot{\phi}, \dot{z})=(0,0)$.
Spin squeezing takes place in the vicinity of unstable fixed points. We are interested in the fixed point located at $z=0$ and $\phi = \frac{\pi}{2}$ according to the location of our initial spin coherent state. The classification of fixed points can be found by analysis of the eigenproblem of the stability matrix ${\cal M}$ which in our case is
\begin{equation}
   {\cal M} = \begin{bmatrix}
\frac{\partial^2\epsilon}{\partial z\partial \phi} & \frac{\partial^2\epsilon}{\partial^2 z} \\
-\frac{\partial^2\epsilon}{\partial^2 \phi} & -\frac{\partial^2\epsilon}{\partial z\partial \phi} 
\end{bmatrix} = \frac{N}{2}\begin{bmatrix}
0 & 1 \\
\eta & 0 
\end{bmatrix} .
\end{equation}
When $\eta\ne0$ then the matrix ${\cal M}$ has two non-degenerate real eigenvalues of the oposite sign $(\lambda_1, \lambda_2) = \frac{1}{2}\sqrt{\eta}N(-1,1)$ and two real eigenvectors
\begin{equation}
   v_1 = 
\begin{bmatrix}
    -\frac{1}{\sqrt{\eta}}     \\
    1     
\end{bmatrix},
v_2 = 
\begin{bmatrix}
    \frac{1}{\sqrt{\eta}}       \\
   1  
\end{bmatrix}. 
\end{equation}
Scalar product of the two eigenvectors $\{v_1, v_2\}$ defines
the angle $\theta$ between in-going and out-going trajectories crossing at the centre of the unstable saddle fixed points 
\begin{equation}\label{eqn:theta}
    \theta = \arccos{ \frac{\langle v_1|v_2\rangle}{|v_1| |v_2|}}.
\end{equation}

In Fig.~\ref{fig:fig2} we show examples of the mean-field phase portraits, i.e. the constant energy lines for $\eta =  \{10^{-5}, 0.04, 0.4, 1\}$. 
The arrows indicate the direction of the evolution, and visualize the dynamics in the vicinity of the fixed point. For $\eta \approx 0 $ the angle between in-going and out-going trajectories is $\theta \approx \pi$ and corresponds to the pure OAT dynamics with non-isolated unstable fixed point. For $\eta>0$ the nature of fixed point changes to the unstable saddle fixed point, see panels (b)-(d), which in the limiting case $\eta=1$ corresponds to the TACT dynamics, panel (d). Note, the angle $\theta$ is approximately $\pi/2$ when the value of anisotropy parameter $\eta$ is one.

\section{Scaling with the system size}\label{sec:Scaling}

In this paragraph we study the scaling of the best squeezing for the anisotropic TACT model~(\ref{eqn:2m}). 
We apply the Gaussian approach within the Bogoliubov-Born-
Green-Kirkwood-Yvon (BBGKY) hierarchy~\cite{PhysRevA.64.013605, PhysRevA.65.053819} which was used in ~\cite{PhysRevA.92.013623} to explain the scaling for the TACT model. Here, we generalize the theory taking into account the values of parameter $\eta$ different than one.

We start with equations of motion for expectation values of spin operators $\langle \dot{\hat{S}}_j\rangle$ which involve terms that depend on
the first-order moments $\langle \hat{S}_j\rangle$ and second-order moments $\langle \hat{S}_i \hat{S}_j\rangle$. Subsequently, the time evolution of the second-order
moments depends on themselves and on third-order moments, and so on. It leads to the BBGKY hierarchy of equations
of motion for expectation values of operator products. The hierarchy is then truncated by keeping the first- and the
second-order moments,
 \begin{align}
      \label{eq:approx}
     \langle \hat{S}_i \hat{S}_j \hat{S}_k\rangle 
     &\simeq 
     \langle \hat{S}_i \hat{S}_j\rangle \langle \hat{S}_k\rangle
     +\langle \hat{S}_j \hat{S}_k\rangle \langle \hat{S}_i\rangle
     + \langle \hat{S}_j \hat{S}_i\rangle \langle \hat{S}_j\rangle\nonumber \\
    & - \langle \hat{S}_i \rangle \langle \hat{S}_j\rangle \langle \hat{S}_k\rangle.
 \end{align}

To perform the scaling analysis, we first introduce a small parameter $\epsilon=1/N$, and transform the spin components into $\hat{J}_j = \sqrt{\epsilon}\hat{S}_j$ which obey cyclic commutation relations $[\hat{J}_x, \hat{J}_y ] = i \sqrt{\epsilon} \hat{J}_z$. The  Hamiltonian (\ref{eqn:2m}) then reads $\hat{H}=\frac{\hbar \chi}{\epsilon} \left( \hat{J}_z^2 - \eta \hat{J}_x^2\right)$. Equations of motion for expectation values of the spin operators $\langle \hat{J}_j \rangle \equiv h_j$,
second order moments $\langle \hat{J}_i\hat{J}_j \rangle\equiv\Delta_{i j }$ and $\langle \hat{J}_j^2 \rangle-\langle \hat{J}_j \rangle^2\equiv\delta_{j }$ read
\begin{align}
    \label{eq:20}
    \dot{h}_y &= 2 (1 + \eta) \Delta_{xz},\\
    \label{eq:21}
    \dot{\Delta}_{xz} &= -2 (\delta_z + \eta \delta_x) h_y ,\\
    \label{eq:22}
    \dot{\delta}_z &= -4 \eta \Delta_{xz} h_y, \\
    \label{eq:23}
    \dot{\delta}_x &= - 4 \Delta_{xz} h_y,
\end{align}
where time is measured in dimensioneless unit $\tau = \chi t/\sqrt{\epsilon}$. 
The initial coherent state at the unstable saddle
fixed point, $|\Psi(0)\rangle=|\theta=\pi/2, \varphi=\pi/2 \rangle$, gives the following initial conditions:
$h_y(0)=(2 \sqrt{\epsilon})^{-1}$, $\delta_z(0)=\delta_x(0)=1/4$ and $\Delta_{xz}(0)=0$. 
In order to find the approximate solution we introduce the two quadratures:
$X=\delta_z + \sqrt{\eta} \Delta_{xz}$ and $Y=\delta_z - \sqrt{\eta} \Delta_{xz}$ obeing the dynamical equations
$\dot{X}=-4 \sqrt{\eta} X h_y$ and $\dot{Y}=-4 \sqrt{\eta} Y h_y$ which have the following solutions:
\begin{align}
    X(t)=X(0) e^{-4 \sqrt{\eta} f(\tau)}, \\
    Y(t)=Y(0) e^{4 \sqrt{\eta} f(\tau)},
\end{align}
where $f(\tau) = \int_0^\tau h_{y}(\tau') d\tau'$ for $\eta\ne 0$. This gives
\begin{align}
    \delta_z(\tau) &= \delta_z(0) \cosh{\left[4 \sqrt{\eta} f(\tau)\right]} \label{eq:deltaz},\\
    \Delta_{xz} (\tau) & = -\frac{\delta_z(0)}{\sqrt{\eta}} 
    \sinh{\left[4 \sqrt{\eta} f(\tau)\right]}\label{eq:Deltaxz},\\
    h_y(\tau)-h_y(0) &= -\frac{\delta_z(0)}{\sqrt{\eta}} 
    \int_0^\tau \sinh{\left[4 \sqrt{\eta} f(\tau')\right]}d \tau'.
\end{align}

In principle, the solution for $h_y$ can be find in self-consistent way, here however, we approximate it by taking the first iteration, namely $f(\tau)\simeq f(0) + f'(0) \tau$, which results in
\begin{equation}
    h_y(\tau) =\frac{1}{2 \sqrt{\epsilon}}
    \left[
    1+\frac{(1 + \eta)\epsilon}{2 \eta}
    \left(
    1 - \cosh(2 \tau \sqrt{\eta/\epsilon})
    \right)
    \right].\label{eq:hy}
\end{equation}
\begin{figure}
    \centering
    \begin{picture}(120, 255)
    \put(-46,-7){\includegraphics[width=.89\linewidth]{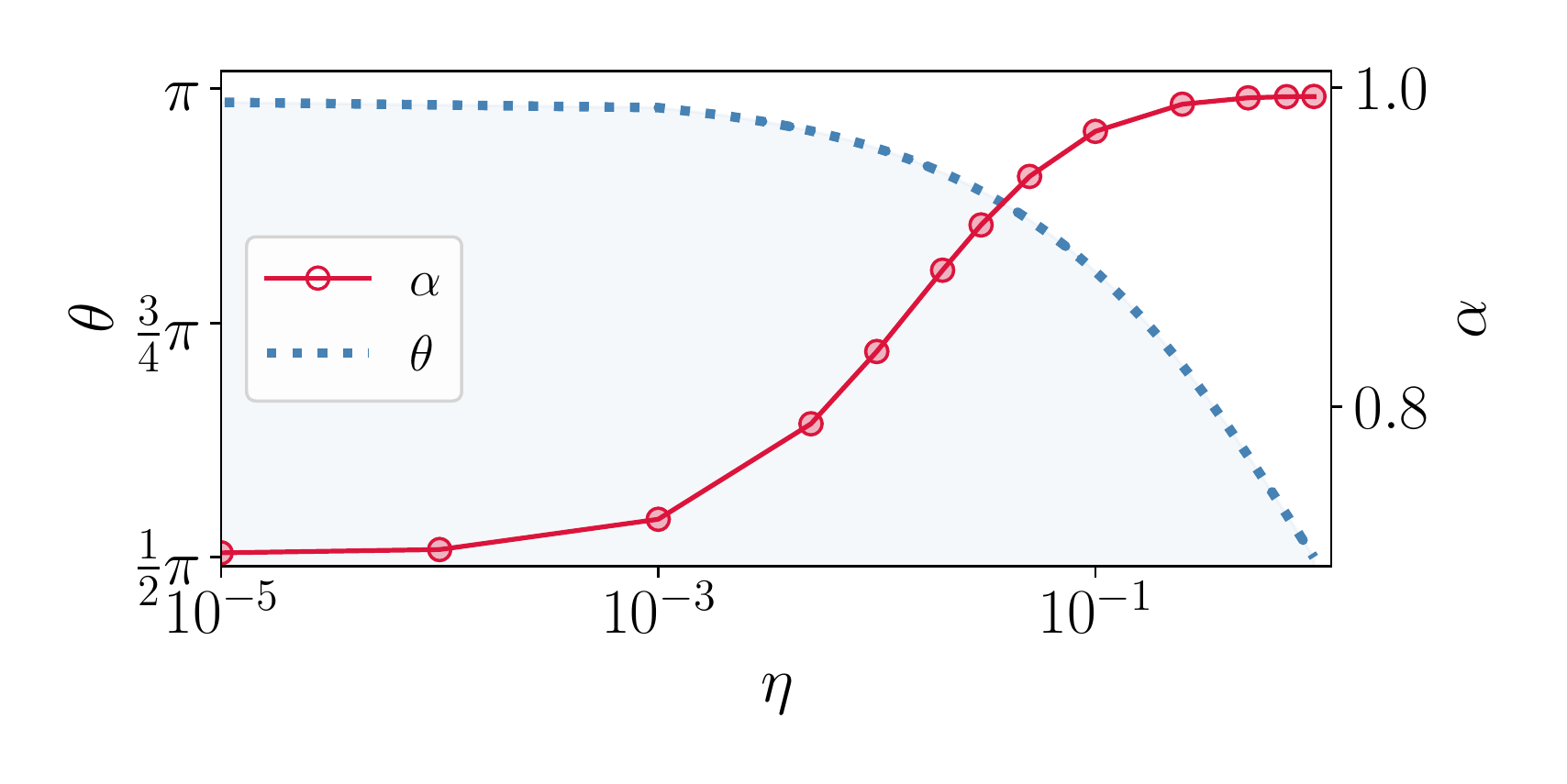}}
    \put(-50,100){\includegraphics[width=.80\linewidth]{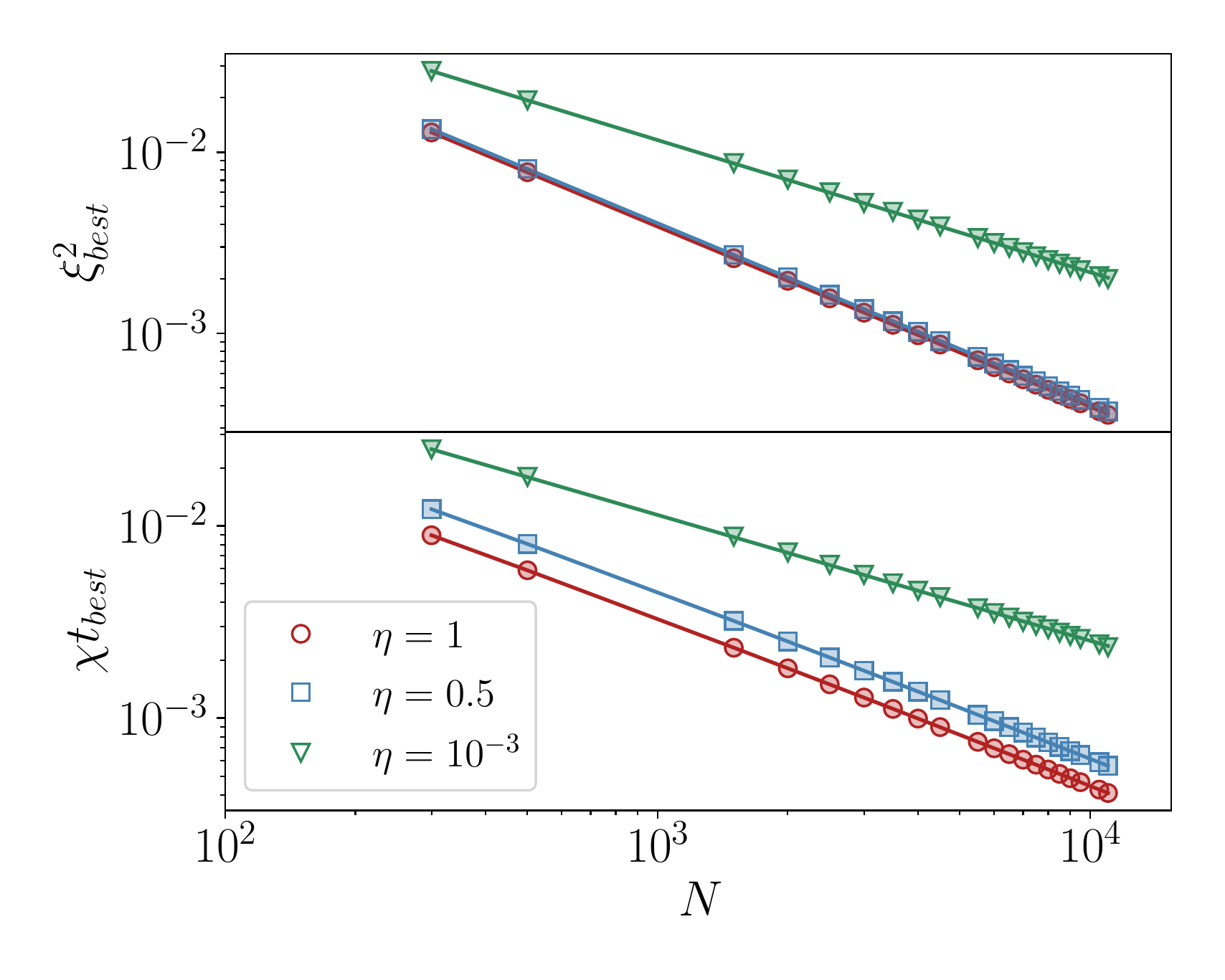}}
    \put(-50,240){(a)}
    \put(-50,85){(b)}
    \end{picture}
  \caption{(a) Examples of the scaling of the best squeezing $\xi^2_{\rm best}$ and the best squeezing time $\chi t_{best}$ with the number of atoms $N$ for various values of the parameter of anisotropy $\eta = \{1,\, 0.5, \, 10^{-3} \}$. (b) Scaling exponents $\alpha$ obtained by fitting the function $\propto N^{-\alpha}$ to the best squeezing $\xi^2_{\rm best}$ is shown by the red points (red solid line is added to guide the eye). The angle $\theta$ between the in-going and out-going mean-field trajectories (see main text) is shown by the dashed blue line. 
  }
  \label{fig:fig3}
\end{figure}
Next, one evaluates the evolution of \eqref{eq:deltaz} and \eqref{eq:Deltaxz} by taking \eqref{eq:hy} in $f(\tau)$. Finally, noting that the spin squeezing parameter \eqref{eq:ssqparameter} is determined by the quadrature $X(t)$, namely $\xi^2 \approx X(t)$, when approximating $\langle S\rangle \approx h_y(0)/\sqrt{\epsilon}$, we obtain the scaling of the best squeezing and the best squeezing time with $N$ by keeping the leading order terms in $\epsilon$, what 
\begin{equation}\label{eq:scaling-ani-TACT}
    \xi^2_{\rm best}\sim 1/N , \,\,\,\,\, \chi t_{\rm best } \sim \frac{\ln ( \eta N)}{ \sqrt{\eta} N},
\end{equation}
when $\eta$ is of the order of one.
We compared the above analytical predictions with the numerically solved set of differential equations \eqref{eq:20}-\eqref{eq:23} and confirmed the scaling~(\ref{eq:scaling-ani-TACT}) when $\eta\in (0.3, 1)$.

The quantitative illustration of the above results can be provided by analyzing the scaling of the best squeezing with $N$ obtained from the numerical time evolution of the TMM Hamiltonian \eqref{eqn:2m}.
Fig.~\ref{fig:fig3} (a) presents the best squeezing $\xi^2_{\rm best}$ and the best squeezing time $\chi t_{\rm best}$ as a function of particle number $N$. Power-law behaviour can be noticed for various $\eta$. Therefore, for each value of the anisotropy parameter $\eta$ we extracted the scaling exponent $\alpha$ by fitting $\xi^2_{\rm best} \sim N^{- \alpha}$. Panel~(b)~of~Fig.~\ref{fig:fig3} shows the change of the fitted exponent $\alpha$ as the function of anisotropy parameter $\eta$ and is compared to the variation of the angle $\theta$. %
A characteristic feature is a change in the value of $\alpha$ when $\eta \in(  10^{-3}, 10^{-1})$.
In the same range $\theta$ diminishes from $\pi$ to approximately $\pi/2$. We conclude, the variation of $\alpha$ is driven by the change in the structure of the unstable fixed point. It is worth to mention here, that $\alpha\approx 1$  when $\eta\approx 1$.
Our results show that the Heisenberg limited level of squeezing is possible in the anisotropic TACT model.

\section{Conclusions}\label{sec:Conclusions}

In this work, we show how the OAT mechanism, generating many-body entanglement, can be accelerated by the long-range interactions via activation of the anisotropic TACT mechanisms. We explain the activation of the TACT
mechanism during competition of contact and dipolar
interactions between bosons in a superfluid phase. We
propose the feasible experimentally quantum simulator
for the amisotropic TACT dynamics based on dipolar two-component
Bose-Hubbard in a one-dimensional optical lattice. With
the help of the scaling analysis, we show that it is possible to obtain a Heisenberg limited level
of squeezing for a weak anisotropy. The anisotropic TACT model accelerates the spin squeezing dynamics compared to OAT
with the improvement of the level of squeezing. Our
protocol allows for fast generation of many-body entangled states with entanglement depth larger than in a standard OAT scenario.

Our work provides an essential step toward generating
many-body entangled states during two-axis counter-twisting protocol in state-of-the-art experimental setups, paving the way for obtaining the Heisenberg limit of spectroscopic measurements in ultracold atoms systems.

\section*{ACKNOWLEDGMENTS}
We gratefully acknowledge discussions with  B.~B.~Laburthe-Tolra. This work was supported by 
the Polish National Science Centre projects
DEC-2019/35/O/ST2/01873 (T.H.Y.), 
DEC-2020/38/L/ST2/00375 (M.D.),
and Grant No. 2019/32/Z/ST2/00016 through the project MAQS under QuantERA, which has received funding from the European Union’s Horizon
2020 research and innovation program under grant agreement no 731473 (E.W. and M.G.).
M.P. acknowledges the support of the Polish National Agency for Academic Exchange, the Bekker programme no: PPN/BEK/2020/1/00317.
ICFO group acknowledges support from: ERC AdG NOQIA; Ministerio de Ciencia y Innovation Agencia Estatal de Investigaciones (PGC2018-097027-B-I00/10.13039/501100011033, CEX2019-000910-S/10.13039/501100011033, Plan National FIDEUA PID2019-106901GB-I00, FPI, QUANTERA MAQS PCI2019-111828-2, QUANTERA DYNAMITE PCI2022-132919, Proyectos de I+D+I “Retos Colaboración” QUSPIN RTC2019-007196-7); European Union NextGenerationEU (PRTR); Fundació Cellex; Fundació Mir-Puig; Generalitat de Catalunya (European Social Fund FEDER and CERCA program (AGAUR Grant No. 2017 SGR 134, QuantumCAT \& U16-011424, co-funded by ERDF Operational Program of Catalonia 2014-2020); Barcelona Supercomputing Center MareNostrum (FI-2022-1-0042); EU Horizon 2020 FET-OPEN OPTOlogic (Grant No 899794); National Science Centre, Poland (Symfonia Grant No. 2016/20/W/ST4/00314); European Union’s Horizon 2020 research and innovation programme under the Marie-Skłodowska-Curie grant agreement No 101029393 (STREDCH) and No 847648 (“La Caixa” Junior Leaders fellowships ID100010434: LCF/BQ/PI19/11690013, LCF/BQ/PI20/11760031, LCF/BQ/PR20/11770012, LCF/BQ/PR21/11840013).
A part of computations were carried out at the Centre of Informatics Tricity Academic Supercomputer \& Network.

\appendix

\section{Dipolar interaction}\label{app:derivation-dipol}

To obtain the lattice version of dipolar interaction one starts with the Hamiltonian (\ref{eq:Hdipgeneral}) and (\ref{eq:dip_int_base}), associates the dipole moment with the Pauli matrices as $\boldsymbol{\mu}_{1} = - \gamma \boldsymbol{\sigma}_{1}$ with $\boldsymbol{\sigma}_1 = (\sigma_x, \sigma_y,\sigma_z)$, and the same at ${\bf r}_2$.
Then one obtains
\begin{align}
 &\hat{H}_{\rm d} =\int d^3 \boldsymbol{r}_1\int d^3 \boldsymbol{r}_2
 \frac{\gamma^2}{|\boldsymbol{r}_1 - \boldsymbol{r}_2|^3} \nonumber \\
 &\times
\left[
(1-3\cos^2\theta_{12})
\left(\hat{J}^z_1\hat{J}^z_2
-
\frac{\hat{J}^+_1\hat{J}^-_2+\hat{J}^-_1\hat{J}^+_2}{4}\right) \right. \nonumber \\
&
- \frac{3}{4} \sin^2\theta_{12} \left(
e^{2i\phi_{12}}\hat{J}^-_1\hat{J}^-_2 + {\rm h.c.}
\right) \nonumber \\
&
\left.
-\frac{3}{4}
\sin2\theta_{12}
\left(
e^{i\phi_{12}}(\hat{J}^z_1\hat{J}^-_2+\hat{J}^-_1\hat{J}^z_2) + {\rm h.c.}
\right) 
\right],
\label{eq:dipolgeneral}
\end{align}
with ${\bf r}_1\ne{\bf r}_2$, and where
$\hat{J}^+_1=\hat{\Psi}_\uparrow^\dagger({\bf r}_1) \hat{\Psi}_\downarrow({\bf r}_1)$,
$\hat{J}^-_1=\hat{\Psi}_\downarrow^\dagger({\bf r}_1) \hat{\Psi}_\uparrow({\bf r}_1)$,
$\hat{J}^z_1=(\hat{\Psi}^\dagger_\uparrow({\bf r}_1) \hat{\Psi}_\uparrow({\bf r}_1) - \hat{\Psi}_\downarrow^\dagger({\bf r}_1) \hat{\Psi}_\downarrow({\bf r}_1))/2$, and similarly at ${\bf r}_2$.
The two angles $\phi,\, \theta$ parameterize the normal vector along ${\bf r}_1 - {\bf r}_2$ direction, namely $\vec{n}_{12}=\frac{{\bf r}_1 - {\bf r}_2}{|{\bf r}_1 - {\bf r}_2|}=(\cos\phi_{12}\sin\theta_{12}, \sin\phi_{12}\sin\theta_{12}, \cos\theta_{12})$.

\begin{figure}
	\centering
	\includegraphics[width=\linewidth]{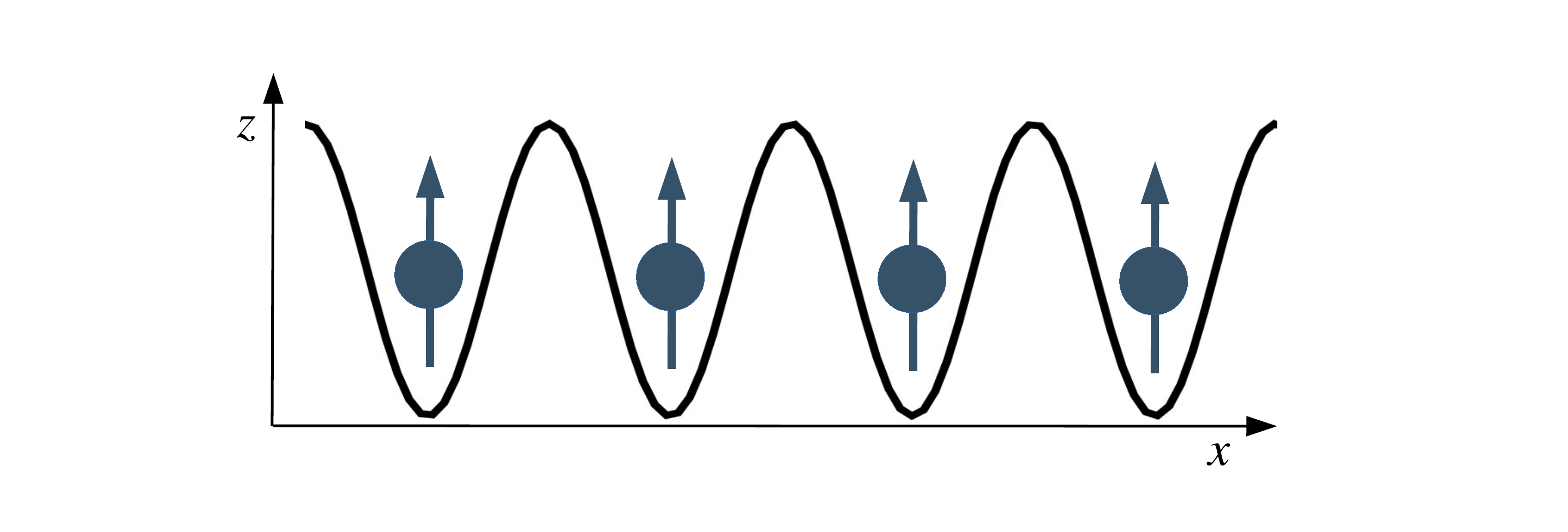}
	\caption{Schematic of the geometry of the system: grey arrows indicates initial configuration of the elementary dipols and the green curve the optical lattice potential.}
	\label{fig:fig_app}
\end{figure}

We assume the system is loaded into one-dimensional optical lattice potential $V_{\rm latt}= V_0 \sin^2 ({2 \pi x/\lambda_{\rm latt}})$, $\lambda_{\rm latt}$ is the lattice wave-length, while remains in its ground state in transverse directions. We assume also the atomic gas is polarized initially along the $z$-axis and the polarization axis sets the quantization axis, as illustrated in Fig.~\ref{fig:fig_app}. Therefore, we consider the following form of the field operator
\begin{align}
    \hat{\Psi}_\uparrow({\bf r})&=\hat{\Phi}_\uparrow(x)\phi(y)\phi(z), \label{eq:factorization}
\end{align}
and we expand $\hat{\Phi}_\uparrow(x)$ in the basis of Wannier functions $w(x-x_j)$ localized around lattice sites, where $x_j$
denotes position of the j-th site in the lowest energy band,
\begin{equation}
    \hat{\Phi}_\uparrow(x) = \sum_j \hat{a}_{\uparrow,j} w(x-x_j),
\end{equation}
where $\hat{a}_{j, \uparrow}$ annihilates an atom in the single-particle Wan-
nier state $w(x-x_j)$ of the lowest energy band localized on the j-th site, in the internal state $\uparrow$.
In (\ref{eq:factorization}) we assume $\phi(y)$ and $\phi(z)$ are the ground state wave-functions of the system in the $y$ and $z$ directions. The same applies for the $\downarrow$ operator.

The geometry of the system, we have chosen, determines the  normal vector, $\vec{n}=(1,0,0)$, and sets the value of $\theta_{12}=\pi/2$ and $\phi_{12}=0$. Taking this into account, in the tight-binding limit when the lattice height is larger than the recoil energy $E_R = (2 \pi)^2/(2 m \lambda_{\rm latt}^2)$ and the Wannier functions are well localized around each lattice site, the dipolar Hamiltonian reduces to
\begin{equation}\label{eq:dipollattice}
	\hat{H}_\mathrm{d} = \sum_{j,k\ne j}
	\frac{\gamma^2}{d^3\left| j-k \right|^3 } 
	\left( \hat{S}_{z,j} \hat{S}_{z,k} -2\hat{S}_{x,j} \hat{S}_{x,k}  +\hat{S}_{y,j} \hat{S}_{y,k} \right),
\end{equation}
due to normalization of the wave functions, and were $d=\lambda_{\rm latt}/2$ will be absorbed in the parameter $\gamma^2$ in the main part of the paper.

It is worth commenting here about the importance of the geometry chosen. There is a symmetry between the $x$ and $y$ axis, i.e., if the lattice would be along the $y$ axis the resulting Hamiltonian (\ref{eq:dipollattice}) would have the factor minus two in the front of the $\hat{S}^y_j \hat{S}^y_k$ term. 
On the other hand, if the lattice would be along $z$ axis, the factor $-2$ appears in the front of $\hat{S}^z_j \hat{S}^z_k$. This has an important consequence in the resulting effective model~(\ref{eqn:2m}) which would be the OAT one.

\section{Numerical evaluation of spin squeezing parameter}\label{app:TMMnumerics}

\subsection{Dipolar Bose-Hubbard model}

We performed the full many-body numerical simulations of $\hat{H}_{\rm dBH}=\hat{H}_{\rm BH}+\hat{H}_{\rm d}$ with \eqref{eq:full_ham} and \eqref{eq:dipolgeneralM}. 
To this end we constructed the Fock states basis, as described in \cite{Plodzien2020}. We implemented numerically the matrix representations of the Hamiltonian $\hat{H}_{\rm dBH}$, and
the initial spin coherent state is
\begin{equation}
    |\Psi(0)\rangle =|\theta, \psi\rangle = e^{-i\hat{S}_z \psi}
    e^{-i\hat{S}_y \theta}
    |\Psi_a\rangle,
\end{equation}
where $|\Psi_a\rangle$ is the ground state of the system when all atoms are in the $|\uparrow \rangle$ state.
The system evolves according to the unitary operator, namely
\begin{equation}
    |\Psi(t)\rangle = e^{-i \hat{H}_{\rm dBH}t/\hbar}|\Psi(0)\rangle,
\end{equation}
and the spin squeezing parameter \eqref{eq:ssqparameter} is calculated.

\subsection{Two-mode model}
In order to find the scaling exponents we perform numerical time evolution of the two-mode model. We express Hamitlonian~\eqref{eqn:2m} in the Fock state basis consisting the vectors of the form $|n, N-n\rangle$,
where $N$ is the total number of atoms, $n$ is the number of the particles in the $|\uparrow\rangle $ state and $N-n$ is the number of the particles in the $|\downarrow\rangle$ state. 


Our initial state is the spin coherent state, which  we obtain as a double rotation of the state $|N,0\rangle$ according to:
\begin{equation}
    |\Psi(0)\rangle =|\theta, \psi\rangle = e^{-i\hat{S}_z \psi}
    e^{-i\hat{S}_y \theta}
    |N, 0\rangle.
\end{equation}
Next, we apply the unitary evolution \begin{equation}
    |\Psi(t)\rangle = e^{-i \hat{H}_{\rm TMM}t/\hbar}|\Psi(0)\rangle,
\end{equation}
and calculate the spin squeezing parameter~\eqref{eq:ssqparameter} and find its first minimum $\xi^2_{\rm best}$, as well as the time at which it occurs $\chi t_{\rm best}$.

\bibliography{reference}

\begin{thebibliography}{57}%
\makeatletter
\providecommand \@ifxundefined [1]{%
 \@ifx{#1\undefined}
}%
\providecommand \@ifnum [1]{%
 \ifnum #1\expandafter \@firstoftwo
 \else \expandafter \@secondoftwo
 \fi
}%
\providecommand \@ifx [1]{%
 \ifx #1\expandafter \@firstoftwo
 \else \expandafter \@secondoftwo
 \fi
}%
\providecommand \natexlab [1]{#1}%
\providecommand \enquote  [1]{``#1''}%
\providecommand \bibnamefont  [1]{#1}%
\providecommand \bibfnamefont [1]{#1}%
\providecommand \citenamefont [1]{#1}%
\providecommand \href@noop [0]{\@secondoftwo}%
\providecommand \href [0]{\begingroup \@sanitize@url \@href}%
\providecommand \@href[1]{\@@startlink{#1}\@@href}%
\providecommand \@@href[1]{\endgroup#1\@@endlink}%
\providecommand \@sanitize@url [0]{\catcode `\\12\catcode `\$12\catcode
  `\&12\catcode `\#12\catcode `\^12\catcode `\_12\catcode `\%12\relax}%
\providecommand \@@startlink[1]{}%
\providecommand \@@endlink[0]{}%
\providecommand \url  [0]{\begingroup\@sanitize@url \@url }%
\providecommand \@url [1]{\endgroup\@href {#1}{\urlprefix }}%
\providecommand \urlprefix  [0]{URL }%
\providecommand \Eprint [0]{\href }%
\providecommand \doibase [0]{http://dx.doi.org/}%
\providecommand \selectlanguage [0]{\@gobble}%
\providecommand \bibinfo  [0]{\@secondoftwo}%
\providecommand \bibfield  [0]{\@secondoftwo}%
\providecommand \translation [1]{[#1]}%
\providecommand \BibitemOpen [0]{}%
\providecommand \bibitemStop [0]{}%
\providecommand \bibitemNoStop [0]{.\EOS\space}%
\providecommand \EOS [0]{\spacefactor3000\relax}%
\providecommand \BibitemShut  [1]{\csname bibitem#1\endcsname}%
\let\auto@bib@innerbib\@empty
\bibitem [{\citenamefont {Ac{\'{\i}}n}\ \emph {et~al.}(2018)\citenamefont
  {Ac{\'{\i}}n}, \citenamefont {Bloch}, \citenamefont {Buhrman}, \citenamefont
  {Calarco}, \citenamefont {Eichler}, \citenamefont {Eisert}, \citenamefont
  {Esteve}, \citenamefont {Gisin}, \citenamefont {Glaser}, \citenamefont
  {Jelezko}, \citenamefont {Kuhr}, \citenamefont {Lewenstein}, \citenamefont
  {Riedel}, \citenamefont {Schmidt}, \citenamefont {Thew}, \citenamefont
  {Wallraff}, \citenamefont {Walmsley},\ and\ \citenamefont
  {Wilhelm}}]{Acin_2018}%
  \BibitemOpen
  \bibfield  {author} {\bibinfo {author} {\bibfnamefont {A.}~\bibnamefont
  {Ac{\'{\i}}n}}, \bibinfo {author} {\bibfnamefont {I.}~\bibnamefont {Bloch}},
  \bibinfo {author} {\bibfnamefont {H.}~\bibnamefont {Buhrman}}, \bibinfo
  {author} {\bibfnamefont {T.}~\bibnamefont {Calarco}}, \bibinfo {author}
  {\bibfnamefont {C.}~\bibnamefont {Eichler}}, \bibinfo {author} {\bibfnamefont
  {J.}~\bibnamefont {Eisert}}, \bibinfo {author} {\bibfnamefont
  {D.}~\bibnamefont {Esteve}}, \bibinfo {author} {\bibfnamefont
  {N.}~\bibnamefont {Gisin}}, \bibinfo {author} {\bibfnamefont {S.~J.}\
  \bibnamefont {Glaser}}, \bibinfo {author} {\bibfnamefont {F.}~\bibnamefont
  {Jelezko}}, \bibinfo {author} {\bibfnamefont {S.}~\bibnamefont {Kuhr}},
  \bibinfo {author} {\bibfnamefont {M.}~\bibnamefont {Lewenstein}}, \bibinfo
  {author} {\bibfnamefont {M.~F.}\ \bibnamefont {Riedel}}, \bibinfo {author}
  {\bibfnamefont {P.~O.}\ \bibnamefont {Schmidt}}, \bibinfo {author}
  {\bibfnamefont {R.}~\bibnamefont {Thew}}, \bibinfo {author} {\bibfnamefont
  {A.}~\bibnamefont {Wallraff}}, \bibinfo {author} {\bibfnamefont
  {I.}~\bibnamefont {Walmsley}}, \ and\ \bibinfo {author} {\bibfnamefont
  {F.~K.}\ \bibnamefont {Wilhelm}},\ }\href {\doibase 10.1088/1367-2630/aad1ea}
  {\bibfield  {journal} {\bibinfo  {journal} {New Journal of Physics}\ }\textbf
  {\bibinfo {volume} {20}},\ \bibinfo {pages} {080201} (\bibinfo {year}
  {2018})}\BibitemShut {NoStop}%
\bibitem [{\citenamefont {Eisert}\ \emph {et~al.}(2020)\citenamefont {Eisert},
  \citenamefont {Hangleiter}, \citenamefont {Walk}, \citenamefont {Roth},
  \citenamefont {Markham}, \citenamefont {Parekh}, \citenamefont {Chabaud},\
  and\ \citenamefont {Kashefi}}]{Eisert2020}%
  \BibitemOpen
  \bibfield  {author} {\bibinfo {author} {\bibfnamefont {J.}~\bibnamefont
  {Eisert}}, \bibinfo {author} {\bibfnamefont {D.}~\bibnamefont {Hangleiter}},
  \bibinfo {author} {\bibfnamefont {N.}~\bibnamefont {Walk}}, \bibinfo {author}
  {\bibfnamefont {I.}~\bibnamefont {Roth}}, \bibinfo {author} {\bibfnamefont
  {D.}~\bibnamefont {Markham}}, \bibinfo {author} {\bibfnamefont
  {R.}~\bibnamefont {Parekh}}, \bibinfo {author} {\bibfnamefont
  {U.}~\bibnamefont {Chabaud}}, \ and\ \bibinfo {author} {\bibfnamefont
  {E.}~\bibnamefont {Kashefi}},\ }\href {\doibase 10.1038/s42254-020-0186-4}
  {\bibfield  {journal} {\bibinfo  {journal} {Nature Reviews Physics}\ }\textbf
  {\bibinfo {volume} {2}},\ \bibinfo {pages} {382} (\bibinfo {year}
  {2020})}\BibitemShut {NoStop}%
\bibitem [{\citenamefont {Kinos}\ \emph {et~al.}(2021)\citenamefont {Kinos},
  \citenamefont {Hunger}, \citenamefont {Kolesov}, \citenamefont {Mølmer},
  \citenamefont {de~Riedmatten}, \citenamefont {Goldner}, \citenamefont
  {Tallaire}, \citenamefont {Morvan}, \citenamefont {Berger}, \citenamefont
  {Welinski}, \citenamefont {Karrai}, \citenamefont {Rippe}, \citenamefont
  {Kröll},\ and\ \citenamefont {Walther}}]{Kinos2021}%
  \BibitemOpen
  \bibfield  {author} {\bibinfo {author} {\bibfnamefont {A.}~\bibnamefont
  {Kinos}}, \bibinfo {author} {\bibfnamefont {D.}~\bibnamefont {Hunger}},
  \bibinfo {author} {\bibfnamefont {R.}~\bibnamefont {Kolesov}}, \bibinfo
  {author} {\bibfnamefont {K.}~\bibnamefont {Mølmer}}, \bibinfo {author}
  {\bibfnamefont {H.}~\bibnamefont {de~Riedmatten}}, \bibinfo {author}
  {\bibfnamefont {P.}~\bibnamefont {Goldner}}, \bibinfo {author} {\bibfnamefont
  {A.}~\bibnamefont {Tallaire}}, \bibinfo {author} {\bibfnamefont
  {L.}~\bibnamefont {Morvan}}, \bibinfo {author} {\bibfnamefont
  {P.}~\bibnamefont {Berger}}, \bibinfo {author} {\bibfnamefont
  {S.}~\bibnamefont {Welinski}}, \bibinfo {author} {\bibfnamefont
  {K.}~\bibnamefont {Karrai}}, \bibinfo {author} {\bibfnamefont
  {L.}~\bibnamefont {Rippe}}, \bibinfo {author} {\bibfnamefont
  {S.}~\bibnamefont {Kröll}}, \ and\ \bibinfo {author} {\bibfnamefont
  {A.}~\bibnamefont {Walther}},\ }\href {\doibase 10.48550/ARXIV.2103.15743}
  {\enquote {\bibinfo {title} {Roadmap for rare-earth quantum computing},}\ }
  (\bibinfo {year} {2021})\BibitemShut {NoStop}%
\bibitem [{\citenamefont {Laucht}\ \emph {et~al.}(2021)\citenamefont {Laucht},
  \citenamefont {Hohls}, \citenamefont {Ubbelohde}, \citenamefont
  {Gonzalez-Zalba}, \citenamefont {Reilly}, \citenamefont {Stobbe},
  \citenamefont {Schröder}, \citenamefont {Scarlino}, \citenamefont {Koski},
  \citenamefont {Dzurak}, \citenamefont {Yang}, \citenamefont {Yoneda},
  \citenamefont {Kuemmeth}, \citenamefont {Bluhm}, \citenamefont {Pla},
  \citenamefont {Hill}, \citenamefont {Salfi}, \citenamefont {Oiwa},
  \citenamefont {Muhonen}, \citenamefont {Verhagen}, \citenamefont {LaHaye},
  \citenamefont {Kim}, \citenamefont {Tsen}, \citenamefont {Culcer},
  \citenamefont {Geresdi}, \citenamefont {Mol}, \citenamefont {Mohan},
  \citenamefont {Jain},\ and\ \citenamefont {Baugh}}]{Laucht_2021}%
  \BibitemOpen
  \bibfield  {author} {\bibinfo {author} {\bibfnamefont {A.}~\bibnamefont
  {Laucht}}, \bibinfo {author} {\bibfnamefont {F.}~\bibnamefont {Hohls}},
  \bibinfo {author} {\bibfnamefont {N.}~\bibnamefont {Ubbelohde}}, \bibinfo
  {author} {\bibfnamefont {M.~F.}\ \bibnamefont {Gonzalez-Zalba}}, \bibinfo
  {author} {\bibfnamefont {D.~J.}\ \bibnamefont {Reilly}}, \bibinfo {author}
  {\bibfnamefont {S.}~\bibnamefont {Stobbe}}, \bibinfo {author} {\bibfnamefont
  {T.}~\bibnamefont {Schröder}}, \bibinfo {author} {\bibfnamefont
  {P.}~\bibnamefont {Scarlino}}, \bibinfo {author} {\bibfnamefont {J.~V.}\
  \bibnamefont {Koski}}, \bibinfo {author} {\bibfnamefont {A.}~\bibnamefont
  {Dzurak}}, \bibinfo {author} {\bibfnamefont {C.-H.}\ \bibnamefont {Yang}},
  \bibinfo {author} {\bibfnamefont {J.}~\bibnamefont {Yoneda}}, \bibinfo
  {author} {\bibfnamefont {F.}~\bibnamefont {Kuemmeth}}, \bibinfo {author}
  {\bibfnamefont {H.}~\bibnamefont {Bluhm}}, \bibinfo {author} {\bibfnamefont
  {J.}~\bibnamefont {Pla}}, \bibinfo {author} {\bibfnamefont {C.}~\bibnamefont
  {Hill}}, \bibinfo {author} {\bibfnamefont {J.}~\bibnamefont {Salfi}},
  \bibinfo {author} {\bibfnamefont {A.}~\bibnamefont {Oiwa}}, \bibinfo {author}
  {\bibfnamefont {J.~T.}\ \bibnamefont {Muhonen}}, \bibinfo {author}
  {\bibfnamefont {E.}~\bibnamefont {Verhagen}}, \bibinfo {author}
  {\bibfnamefont {M.~D.}\ \bibnamefont {LaHaye}}, \bibinfo {author}
  {\bibfnamefont {H.~H.}\ \bibnamefont {Kim}}, \bibinfo {author} {\bibfnamefont
  {A.~W.}\ \bibnamefont {Tsen}}, \bibinfo {author} {\bibfnamefont
  {D.}~\bibnamefont {Culcer}}, \bibinfo {author} {\bibfnamefont
  {A.}~\bibnamefont {Geresdi}}, \bibinfo {author} {\bibfnamefont {J.~A.}\
  \bibnamefont {Mol}}, \bibinfo {author} {\bibfnamefont {V.}~\bibnamefont
  {Mohan}}, \bibinfo {author} {\bibfnamefont {P.~K.}\ \bibnamefont {Jain}}, \
  and\ \bibinfo {author} {\bibfnamefont {J.}~\bibnamefont {Baugh}},\ }\href
  {\doibase 10.1088/1361-6528/abb333} {\bibfield  {journal} {\bibinfo
  {journal} {Nanotechnology}\ }\textbf {\bibinfo {volume} {32}},\ \bibinfo
  {pages} {162003} (\bibinfo {year} {2021})}\BibitemShut {NoStop}%
\bibitem [{\citenamefont {Becher}\ \emph {et~al.}(2022)\citenamefont {Becher},
  \citenamefont {Gao}, \citenamefont {Kar}, \citenamefont {Marciniak},
  \citenamefont {Monz}, \citenamefont {Bartholomew}, \citenamefont {Goldner},
  \citenamefont {Loh}, \citenamefont {Marcellina}, \citenamefont {Goh},
  \citenamefont {Koh}, \citenamefont {Weber}, \citenamefont {Mu}, \citenamefont
  {Tsai}, \citenamefont {Yan}, \citenamefont {Gyger}, \citenamefont
  {Steinhauer},\ and\ \citenamefont {Zwiller}}]{Zwiller2022}%
  \BibitemOpen
  \bibfield  {author} {\bibinfo {author} {\bibfnamefont {C.}~\bibnamefont
  {Becher}}, \bibinfo {author} {\bibfnamefont {W.}~\bibnamefont {Gao}},
  \bibinfo {author} {\bibfnamefont {S.}~\bibnamefont {Kar}}, \bibinfo {author}
  {\bibfnamefont {C.}~\bibnamefont {Marciniak}}, \bibinfo {author}
  {\bibfnamefont {T.}~\bibnamefont {Monz}}, \bibinfo {author} {\bibfnamefont
  {J.~G.}\ \bibnamefont {Bartholomew}}, \bibinfo {author} {\bibfnamefont
  {P.}~\bibnamefont {Goldner}}, \bibinfo {author} {\bibfnamefont
  {H.}~\bibnamefont {Loh}}, \bibinfo {author} {\bibfnamefont {E.}~\bibnamefont
  {Marcellina}}, \bibinfo {author} {\bibfnamefont {K.~E.~J.}\ \bibnamefont
  {Goh}}, \bibinfo {author} {\bibfnamefont {T.~S.}\ \bibnamefont {Koh}},
  \bibinfo {author} {\bibfnamefont {B.}~\bibnamefont {Weber}}, \bibinfo
  {author} {\bibfnamefont {Z.}~\bibnamefont {Mu}}, \bibinfo {author}
  {\bibfnamefont {J.-Y.}\ \bibnamefont {Tsai}}, \bibinfo {author}
  {\bibfnamefont {Q.}~\bibnamefont {Yan}}, \bibinfo {author} {\bibfnamefont
  {S.}~\bibnamefont {Gyger}}, \bibinfo {author} {\bibfnamefont
  {S.}~\bibnamefont {Steinhauer}}, \ and\ \bibinfo {author} {\bibfnamefont
  {V.}~\bibnamefont {Zwiller}},\ }\href {\doibase 10.48550/ARXIV.2202.07309}
  {\enquote {\bibinfo {title} {2022 roadmap for materials for quantum
  technologies},}\ } (\bibinfo {year} {2022})\BibitemShut {NoStop}%
\bibitem [{\citenamefont {Fraxanet}\ \emph {et~al.}(2022)\citenamefont
  {Fraxanet}, \citenamefont {Salamon},\ and\ \citenamefont
  {Lewenstein}}]{Fraxanet2022}%
  \BibitemOpen
  \bibfield  {author} {\bibinfo {author} {\bibfnamefont {J.}~\bibnamefont
  {Fraxanet}}, \bibinfo {author} {\bibfnamefont {T.}~\bibnamefont {Salamon}}, \
  and\ \bibinfo {author} {\bibfnamefont {M.}~\bibnamefont {Lewenstein}},\
  }\href {\doibase 10.48550/ARXIV.2204.08905} {\enquote {\bibinfo {title} {The
  coming decades of quantum simulation},}\ } (\bibinfo {year}
  {2022})\BibitemShut {NoStop}%
\bibitem [{\citenamefont {Kitagawa}\ and\ \citenamefont
  {Ueda}(1993)}]{Kitagawa1993}%
  \BibitemOpen
  \bibfield  {author} {\bibinfo {author} {\bibfnamefont {M.}~\bibnamefont
  {Kitagawa}}\ and\ \bibinfo {author} {\bibfnamefont {M.}~\bibnamefont
  {Ueda}},\ }\href {\doibase 10.1103/PhysRevA.47.5138} {\bibfield  {journal}
  {\bibinfo  {journal} {Phys. Rev. A}\ }\textbf {\bibinfo {volume} {47}},\
  \bibinfo {pages} {5138} (\bibinfo {year} {1993})}\BibitemShut {NoStop}%
\bibitem [{\citenamefont {Wineland}\ \emph {et~al.}(1994)\citenamefont
  {Wineland}, \citenamefont {Bollinger}, \citenamefont {Itano},\ and\
  \citenamefont {Heinzen}}]{Wineland1994}%
  \BibitemOpen
  \bibfield  {author} {\bibinfo {author} {\bibfnamefont {D.~J.}\ \bibnamefont
  {Wineland}}, \bibinfo {author} {\bibfnamefont {J.~J.}\ \bibnamefont
  {Bollinger}}, \bibinfo {author} {\bibfnamefont {W.~M.}\ \bibnamefont
  {Itano}}, \ and\ \bibinfo {author} {\bibfnamefont {D.~J.}\ \bibnamefont
  {Heinzen}},\ }\href {\doibase 10.1103/PhysRevA.50.67} {\bibfield  {journal}
  {\bibinfo  {journal} {Phys. Rev. A}\ }\textbf {\bibinfo {volume} {50}},\
  \bibinfo {pages} {67} (\bibinfo {year} {1994})}\BibitemShut {NoStop}%
\bibitem [{\citenamefont {Fadel}\ \emph {et~al.}(2018)\citenamefont {Fadel},
  \citenamefont {Zibold}, \citenamefont {Décamps},\ and\ \citenamefont
  {Treutlein}}]{Fadel2018}%
  \BibitemOpen
  \bibfield  {author} {\bibinfo {author} {\bibfnamefont {M.}~\bibnamefont
  {Fadel}}, \bibinfo {author} {\bibfnamefont {T.}~\bibnamefont {Zibold}},
  \bibinfo {author} {\bibfnamefont {B.}~\bibnamefont {Décamps}}, \ and\
  \bibinfo {author} {\bibfnamefont {P.}~\bibnamefont {Treutlein}},\ }\href
  {\doibase 10.1126/science.aao1850} {\bibfield  {journal} {\bibinfo  {journal}
  {Science}\ }\textbf {\bibinfo {volume} {360}},\ \bibinfo {pages} {409–413}
  (\bibinfo {year} {2018})}\BibitemShut {NoStop}%
\bibitem [{\citenamefont {Evrard}\ \emph {et~al.}(2019)\citenamefont {Evrard},
  \citenamefont {Makhalov}, \citenamefont {Chalopin}, \citenamefont
  {Sidorenkov}, \citenamefont {Dalibard}, \citenamefont {Lopes},\ and\
  \citenamefont {Nascimbene}}]{PhysRevLett.122.173601}%
  \BibitemOpen
  \bibfield  {author} {\bibinfo {author} {\bibfnamefont {A.}~\bibnamefont
  {Evrard}}, \bibinfo {author} {\bibfnamefont {V.}~\bibnamefont {Makhalov}},
  \bibinfo {author} {\bibfnamefont {T.}~\bibnamefont {Chalopin}}, \bibinfo
  {author} {\bibfnamefont {L.~A.}\ \bibnamefont {Sidorenkov}}, \bibinfo
  {author} {\bibfnamefont {J.}~\bibnamefont {Dalibard}}, \bibinfo {author}
  {\bibfnamefont {R.}~\bibnamefont {Lopes}}, \ and\ \bibinfo {author}
  {\bibfnamefont {S.}~\bibnamefont {Nascimbene}},\ }\href {\doibase
  10.1103/PhysRevLett.122.173601} {\bibfield  {journal} {\bibinfo  {journal}
  {Phys. Rev. Lett.}\ }\textbf {\bibinfo {volume} {122}},\ \bibinfo {pages}
  {173601} (\bibinfo {year} {2019})}\BibitemShut {NoStop}%
\bibitem [{\citenamefont {Hosten}\ \emph {et~al.}(2016)\citenamefont {Hosten},
  \citenamefont {Engelsen}, \citenamefont {Krishnakumar},\ and\ \citenamefont
  {Kasevich}}]{Hosten2016}%
  \BibitemOpen
  \bibfield  {author} {\bibinfo {author} {\bibfnamefont {O.}~\bibnamefont
  {Hosten}}, \bibinfo {author} {\bibfnamefont {N.~J.}\ \bibnamefont
  {Engelsen}}, \bibinfo {author} {\bibfnamefont {R.}~\bibnamefont
  {Krishnakumar}}, \ and\ \bibinfo {author} {\bibfnamefont {M.~A.}\
  \bibnamefont {Kasevich}},\ }\href {\doibase 10.1038/nature16176} {\bibfield
  {journal} {\bibinfo  {journal} {Nature}\ }\textbf {\bibinfo {volume} {529}},\
  \bibinfo {pages} {505} (\bibinfo {year} {2016})}\BibitemShut {NoStop}%
\bibitem [{\citenamefont {Pedrozo-Peñafiel}\ \emph {et~al.}(2020)\citenamefont
  {Pedrozo-Peñafiel}, \citenamefont {Colombo}, \citenamefont {Shu},
  \citenamefont {Adiyatullin}, \citenamefont {Li}, \citenamefont {Mendez},
  \citenamefont {Braverman}, \citenamefont {Kawasaki}, \citenamefont
  {Akamatsu}, \citenamefont {Xiao},\ and\ \citenamefont
  {et~al.}}]{Pedrozo2020}%
  \BibitemOpen
  \bibfield  {author} {\bibinfo {author} {\bibfnamefont {E.}~\bibnamefont
  {Pedrozo-Peñafiel}}, \bibinfo {author} {\bibfnamefont {S.}~\bibnamefont
  {Colombo}}, \bibinfo {author} {\bibfnamefont {C.}~\bibnamefont {Shu}},
  \bibinfo {author} {\bibfnamefont {A.~F.}\ \bibnamefont {Adiyatullin}},
  \bibinfo {author} {\bibfnamefont {Z.}~\bibnamefont {Li}}, \bibinfo {author}
  {\bibfnamefont {E.}~\bibnamefont {Mendez}}, \bibinfo {author} {\bibfnamefont
  {B.}~\bibnamefont {Braverman}}, \bibinfo {author} {\bibfnamefont
  {A.}~\bibnamefont {Kawasaki}}, \bibinfo {author} {\bibfnamefont
  {D.}~\bibnamefont {Akamatsu}}, \bibinfo {author} {\bibfnamefont
  {Y.}~\bibnamefont {Xiao}}, \ and\ \bibinfo {author} {\bibnamefont {et~al.}},\
  }\href {\doibase 10.1038/s41586-020-3006-1} {\bibfield  {journal} {\bibinfo
  {journal} {Nature}\ }\textbf {\bibinfo {volume} {588}},\ \bibinfo {pages}
  {414–418} (\bibinfo {year} {2020})}\BibitemShut {NoStop}%
\bibitem [{\citenamefont {Bao}\ \emph {et~al.}(2020)\citenamefont {Bao},
  \citenamefont {Duan}, \citenamefont {Jin}, \citenamefont {Lu}, \citenamefont
  {Li}, \citenamefont {Qu}, \citenamefont {Wang}, \citenamefont {Novikova},
  \citenamefont {Mikhailov}, \citenamefont {Zhao}, \citenamefont {M{\o}lmer},
  \citenamefont {Shen},\ and\ \citenamefont {Xiao}}]{Bao2020}%
  \BibitemOpen
  \bibfield  {author} {\bibinfo {author} {\bibfnamefont {H.}~\bibnamefont
  {Bao}}, \bibinfo {author} {\bibfnamefont {J.}~\bibnamefont {Duan}}, \bibinfo
  {author} {\bibfnamefont {S.}~\bibnamefont {Jin}}, \bibinfo {author}
  {\bibfnamefont {X.}~\bibnamefont {Lu}}, \bibinfo {author} {\bibfnamefont
  {P.}~\bibnamefont {Li}}, \bibinfo {author} {\bibfnamefont {W.}~\bibnamefont
  {Qu}}, \bibinfo {author} {\bibfnamefont {M.}~\bibnamefont {Wang}}, \bibinfo
  {author} {\bibfnamefont {I.}~\bibnamefont {Novikova}}, \bibinfo {author}
  {\bibfnamefont {E.~E.}\ \bibnamefont {Mikhailov}}, \bibinfo {author}
  {\bibfnamefont {K.-F.}\ \bibnamefont {Zhao}}, \bibinfo {author}
  {\bibfnamefont {K.}~\bibnamefont {M{\o}lmer}}, \bibinfo {author}
  {\bibfnamefont {H.}~\bibnamefont {Shen}}, \ and\ \bibinfo {author}
  {\bibfnamefont {Y.}~\bibnamefont {Xiao}},\ }\href {\doibase
  10.1038/s41586-020-2243-7} {\bibfield  {journal} {\bibinfo  {journal}
  {Nature}\ }\textbf {\bibinfo {volume} {581}},\ \bibinfo {pages} {159}
  (\bibinfo {year} {2020})}\BibitemShut {NoStop}%
\bibitem [{\citenamefont {Tura}\ \emph {et~al.}(2014)\citenamefont {Tura},
  \citenamefont {Augusiak}, \citenamefont {Sainz}, \citenamefont {V{\'e}rtesi},
  \citenamefont {Lewenstein},\ and\ \citenamefont {Ac{\'\i}n}}]{Tura1256}%
  \BibitemOpen
  \bibfield  {author} {\bibinfo {author} {\bibfnamefont {J.}~\bibnamefont
  {Tura}}, \bibinfo {author} {\bibfnamefont {R.}~\bibnamefont {Augusiak}},
  \bibinfo {author} {\bibfnamefont {A.~B.}\ \bibnamefont {Sainz}}, \bibinfo
  {author} {\bibfnamefont {T.}~\bibnamefont {V{\'e}rtesi}}, \bibinfo {author}
  {\bibfnamefont {M.}~\bibnamefont {Lewenstein}}, \ and\ \bibinfo {author}
  {\bibfnamefont {A.}~\bibnamefont {Ac{\'\i}n}},\ }\href@noop {} {\bibfield
  {journal} {\bibinfo  {journal} {Science}\ }\textbf {\bibinfo {volume}
  {344}},\ \bibinfo {pages} {1256} (\bibinfo {year} {2014})}\BibitemShut
  {NoStop}%
\bibitem [{\citenamefont {Schmied}\ \emph {et~al.}(2016)\citenamefont
  {Schmied}, \citenamefont {Bancal}, \citenamefont {Allard}, \citenamefont
  {Fadel}, \citenamefont {Scarani}, \citenamefont {Treutlein},\ and\
  \citenamefont {Sangouard}}]{schmied2016bell}%
  \BibitemOpen
  \bibfield  {author} {\bibinfo {author} {\bibfnamefont {R.}~\bibnamefont
  {Schmied}}, \bibinfo {author} {\bibfnamefont {J.-D.}\ \bibnamefont {Bancal}},
  \bibinfo {author} {\bibfnamefont {B.}~\bibnamefont {Allard}}, \bibinfo
  {author} {\bibfnamefont {M.}~\bibnamefont {Fadel}}, \bibinfo {author}
  {\bibfnamefont {V.}~\bibnamefont {Scarani}}, \bibinfo {author} {\bibfnamefont
  {P.}~\bibnamefont {Treutlein}}, \ and\ \bibinfo {author} {\bibfnamefont
  {N.}~\bibnamefont {Sangouard}},\ }\href@noop {} {\bibfield  {journal}
  {\bibinfo  {journal} {Science}\ }\textbf {\bibinfo {volume} {352}},\ \bibinfo
  {pages} {441} (\bibinfo {year} {2016})}\BibitemShut {NoStop}%
\bibitem [{\citenamefont {Aloy}\ \emph {et~al.}(2019)\citenamefont {Aloy},
  \citenamefont {Tura}, \citenamefont {Baccari}, \citenamefont {Ac\'{\i}n},
  \citenamefont {Lewenstein},\ and\ \citenamefont {Augusiak}}]{Aloy2019}%
  \BibitemOpen
  \bibfield  {author} {\bibinfo {author} {\bibfnamefont {A.}~\bibnamefont
  {Aloy}}, \bibinfo {author} {\bibfnamefont {J.}~\bibnamefont {Tura}}, \bibinfo
  {author} {\bibfnamefont {F.}~\bibnamefont {Baccari}}, \bibinfo {author}
  {\bibfnamefont {A.}~\bibnamefont {Ac\'{\i}n}}, \bibinfo {author}
  {\bibfnamefont {M.}~\bibnamefont {Lewenstein}}, \ and\ \bibinfo {author}
  {\bibfnamefont {R.}~\bibnamefont {Augusiak}},\ }\href {\doibase
  10.1103/PhysRevLett.123.100507} {\bibfield  {journal} {\bibinfo  {journal}
  {Phys. Rev. Lett.}\ }\textbf {\bibinfo {volume} {123}},\ \bibinfo {pages}
  {100507} (\bibinfo {year} {2019})}\BibitemShut {NoStop}%
\bibitem [{\citenamefont {Baccari}\ \emph {et~al.}(2019)\citenamefont
  {Baccari}, \citenamefont {Tura}, \citenamefont {Fadel}, \citenamefont {Aloy},
  \citenamefont {Bancal}, \citenamefont {Sangouard}, \citenamefont
  {Lewenstein}, \citenamefont {Ac\'{\i}n},\ and\ \citenamefont
  {Augusiak}}]{Baccari2019}%
  \BibitemOpen
  \bibfield  {author} {\bibinfo {author} {\bibfnamefont {F.}~\bibnamefont
  {Baccari}}, \bibinfo {author} {\bibfnamefont {J.}~\bibnamefont {Tura}},
  \bibinfo {author} {\bibfnamefont {M.}~\bibnamefont {Fadel}}, \bibinfo
  {author} {\bibfnamefont {A.}~\bibnamefont {Aloy}}, \bibinfo {author}
  {\bibfnamefont {J.-D.}\ \bibnamefont {Bancal}}, \bibinfo {author}
  {\bibfnamefont {N.}~\bibnamefont {Sangouard}}, \bibinfo {author}
  {\bibfnamefont {M.}~\bibnamefont {Lewenstein}}, \bibinfo {author}
  {\bibfnamefont {A.}~\bibnamefont {Ac\'{\i}n}}, \ and\ \bibinfo {author}
  {\bibfnamefont {R.}~\bibnamefont {Augusiak}},\ }\href {\doibase
  10.1103/PhysRevA.100.022121} {\bibfield  {journal} {\bibinfo  {journal}
  {Phys. Rev. A}\ }\textbf {\bibinfo {volume} {100}},\ \bibinfo {pages}
  {022121} (\bibinfo {year} {2019})}\BibitemShut {NoStop}%
\bibitem [{\citenamefont {Tura}\ \emph {et~al.}(2019)\citenamefont {Tura},
  \citenamefont {Aloy}, \citenamefont {Baccari}, \citenamefont {Ac\'{\i}n},
  \citenamefont {Lewenstein},\ and\ \citenamefont {Augusiak}}]{Tura2019}%
  \BibitemOpen
  \bibfield  {author} {\bibinfo {author} {\bibfnamefont {J.}~\bibnamefont
  {Tura}}, \bibinfo {author} {\bibfnamefont {A.}~\bibnamefont {Aloy}}, \bibinfo
  {author} {\bibfnamefont {F.}~\bibnamefont {Baccari}}, \bibinfo {author}
  {\bibfnamefont {A.}~\bibnamefont {Ac\'{\i}n}}, \bibinfo {author}
  {\bibfnamefont {M.}~\bibnamefont {Lewenstein}}, \ and\ \bibinfo {author}
  {\bibfnamefont {R.}~\bibnamefont {Augusiak}},\ }\href {\doibase
  10.1103/PhysRevA.100.032307} {\bibfield  {journal} {\bibinfo  {journal}
  {Phys. Rev. A}\ }\textbf {\bibinfo {volume} {100}},\ \bibinfo {pages}
  {032307} (\bibinfo {year} {2019})}\BibitemShut {NoStop}%
\bibitem [{\citenamefont {M\"uller-Rigat}\ \emph {et~al.}(2021)\citenamefont
  {M\"uller-Rigat}, \citenamefont {Aloy}, \citenamefont {Lewenstein},\ and\
  \citenamefont {Fr\'erot}}]{PRXQuantum.2.030329}%
  \BibitemOpen
  \bibfield  {author} {\bibinfo {author} {\bibfnamefont {G.}~\bibnamefont
  {M\"uller-Rigat}}, \bibinfo {author} {\bibfnamefont {A.}~\bibnamefont
  {Aloy}}, \bibinfo {author} {\bibfnamefont {M.}~\bibnamefont {Lewenstein}}, \
  and\ \bibinfo {author} {\bibfnamefont {I.}~\bibnamefont {Fr\'erot}},\ }\href
  {\doibase 10.1103/PRXQuantum.2.030329} {\bibfield  {journal} {\bibinfo
  {journal} {PRX Quantum}\ }\textbf {\bibinfo {volume} {2}},\ \bibinfo {pages}
  {030329} (\bibinfo {year} {2021})}\BibitemShut {NoStop}%
\bibitem [{\citenamefont {Płodzień}\ \emph {et~al.}(2022)\citenamefont
  {Płodzień}, \citenamefont {Lewenstein}, \citenamefont {Witkowska},\ and\
  \citenamefont {Chwedeńczuk}}]{Plodzien2022}%
  \BibitemOpen
  \bibfield  {author} {\bibinfo {author} {\bibfnamefont {M.}~\bibnamefont
  {Płodzień}}, \bibinfo {author} {\bibfnamefont {M.}~\bibnamefont
  {Lewenstein}}, \bibinfo {author} {\bibfnamefont {E.}~\bibnamefont
  {Witkowska}}, \ and\ \bibinfo {author} {\bibfnamefont {J.}~\bibnamefont
  {Chwedeńczuk}},\ }\href {\doibase 10.48550/ARXIV.2206.10542} {\  (\bibinfo
  {year} {2022}),\ 10.48550/ARXIV.2206.10542}\BibitemShut {NoStop}%
\bibitem [{\citenamefont {S{\o}rensen}\ \emph {et~al.}(2001)\citenamefont
  {S{\o}rensen}, \citenamefont {Duan}, \citenamefont {Cirac},\ and\
  \citenamefont {Zoller}}]{Sorensen2001_Nature}%
  \BibitemOpen
  \bibfield  {author} {\bibinfo {author} {\bibfnamefont {A.}~\bibnamefont
  {S{\o}rensen}}, \bibinfo {author} {\bibfnamefont {L.-M.}\ \bibnamefont
  {Duan}}, \bibinfo {author} {\bibfnamefont {J.~I.}\ \bibnamefont {Cirac}}, \
  and\ \bibinfo {author} {\bibfnamefont {P.}~\bibnamefont {Zoller}},\ }\href
  {\doibase 10.1038/35051038} {\bibfield  {journal} {\bibinfo  {journal}
  {Nature}\ }\textbf {\bibinfo {volume} {409}},\ \bibinfo {pages} {63}
  (\bibinfo {year} {2001})}\BibitemShut {NoStop}%
\bibitem [{\citenamefont {Pezz\'e}\ and\ \citenamefont
  {Smerzi}(2009)}]{Pezze2009}%
  \BibitemOpen
  \bibfield  {author} {\bibinfo {author} {\bibfnamefont {L.}~\bibnamefont
  {Pezz\'e}}\ and\ \bibinfo {author} {\bibfnamefont {A.}~\bibnamefont
  {Smerzi}},\ }\href {\doibase 10.1103/PhysRevLett.102.100401} {\bibfield
  {journal} {\bibinfo  {journal} {Phys. Rev. Lett.}\ }\textbf {\bibinfo
  {volume} {102}},\ \bibinfo {pages} {100401} (\bibinfo {year}
  {2009})}\BibitemShut {NoStop}%
\bibitem [{\citenamefont {Hyllus}\ \emph {et~al.}(2012)\citenamefont {Hyllus},
  \citenamefont {Laskowski}, \citenamefont {Krischek}, \citenamefont
  {Schwemmer}, \citenamefont {Wieczorek}, \citenamefont {Weinfurter},
  \citenamefont {Pezz\'e},\ and\ \citenamefont {Smerzi}}]{Hyllus2012}%
  \BibitemOpen
  \bibfield  {author} {\bibinfo {author} {\bibfnamefont {P.}~\bibnamefont
  {Hyllus}}, \bibinfo {author} {\bibfnamefont {W.}~\bibnamefont {Laskowski}},
  \bibinfo {author} {\bibfnamefont {R.}~\bibnamefont {Krischek}}, \bibinfo
  {author} {\bibfnamefont {C.}~\bibnamefont {Schwemmer}}, \bibinfo {author}
  {\bibfnamefont {W.}~\bibnamefont {Wieczorek}}, \bibinfo {author}
  {\bibfnamefont {H.}~\bibnamefont {Weinfurter}}, \bibinfo {author}
  {\bibfnamefont {L.}~\bibnamefont {Pezz\'e}}, \ and\ \bibinfo {author}
  {\bibfnamefont {A.}~\bibnamefont {Smerzi}},\ }\href {\doibase
  10.1103/PhysRevA.85.022321} {\bibfield  {journal} {\bibinfo  {journal} {Phys.
  Rev. A}\ }\textbf {\bibinfo {volume} {85}},\ \bibinfo {pages} {022321}
  (\bibinfo {year} {2012})}\BibitemShut {NoStop}%
\bibitem [{\citenamefont {T\'oth}(2012)}]{Toth2012}%
  \BibitemOpen
  \bibfield  {author} {\bibinfo {author} {\bibfnamefont {G.}~\bibnamefont
  {T\'oth}},\ }\href {\doibase 10.1103/PhysRevA.85.022322} {\bibfield
  {journal} {\bibinfo  {journal} {Phys. Rev. A}\ }\textbf {\bibinfo {volume}
  {85}},\ \bibinfo {pages} {022322} (\bibinfo {year} {2012})}\BibitemShut
  {NoStop}%
\bibitem [{\citenamefont {Kajtoch}\ and\ \citenamefont
  {Witkowska}(2015)}]{PhysRevA.92.013623}%
  \BibitemOpen
  \bibfield  {author} {\bibinfo {author} {\bibfnamefont {D.}~\bibnamefont
  {Kajtoch}}\ and\ \bibinfo {author} {\bibfnamefont {E.}~\bibnamefont
  {Witkowska}},\ }\href {\doibase 10.1103/PhysRevA.92.013623} {\bibfield
  {journal} {\bibinfo  {journal} {Phys. Rev. A}\ }\textbf {\bibinfo {volume}
  {92}},\ \bibinfo {pages} {013623} (\bibinfo {year} {2015})}\BibitemShut
  {NoStop}%
\bibitem [{\citenamefont {S\o{}rensen}\ and\ \citenamefont
  {M\o{}lmer}(1999)}]{Sorensen1999PRL}%
  \BibitemOpen
  \bibfield  {author} {\bibinfo {author} {\bibfnamefont {A.}~\bibnamefont
  {S\o{}rensen}}\ and\ \bibinfo {author} {\bibfnamefont {K.}~\bibnamefont
  {M\o{}lmer}},\ }\href {\doibase 10.1103/PhysRevLett.83.2274} {\bibfield
  {journal} {\bibinfo  {journal} {Phys. Rev. Lett.}\ }\textbf {\bibinfo
  {volume} {83}},\ \bibinfo {pages} {2274} (\bibinfo {year}
  {1999})}\BibitemShut {NoStop}%
\bibitem [{\citenamefont {Sørensen}\ \emph {et~al.}(2001)\citenamefont
  {Sørensen}, \citenamefont {Duan}, \citenamefont {Cirac},\ and\ \citenamefont
  {Zoller}}]{Sorensen2001}%
  \BibitemOpen
  \bibfield  {author} {\bibinfo {author} {\bibfnamefont {A.}~\bibnamefont
  {Sørensen}}, \bibinfo {author} {\bibfnamefont {L.-M.}\ \bibnamefont {Duan}},
  \bibinfo {author} {\bibfnamefont {J.~I.}\ \bibnamefont {Cirac}}, \ and\
  \bibinfo {author} {\bibfnamefont {P.}~\bibnamefont {Zoller}},\ }\href
  {\doibase 10.1038/35051038} {\bibfield  {journal} {\bibinfo  {journal}
  {Nature}\ }\textbf {\bibinfo {volume} {409}},\ \bibinfo {pages} {63–66}
  (\bibinfo {year} {2001})}\BibitemShut {NoStop}%
\bibitem [{\citenamefont {Riedel}\ \emph {et~al.}(2010)\citenamefont {Riedel},
  \citenamefont {Böhi}, \citenamefont {Li}, \citenamefont {Hänsch},
  \citenamefont {Sinatra},\ and\ \citenamefont {Treutlein}}]{Treutlein2010}%
  \BibitemOpen
  \bibfield  {author} {\bibinfo {author} {\bibfnamefont {M.~F.}\ \bibnamefont
  {Riedel}}, \bibinfo {author} {\bibfnamefont {P.}~\bibnamefont {Böhi}},
  \bibinfo {author} {\bibfnamefont {Y.}~\bibnamefont {Li}}, \bibinfo {author}
  {\bibfnamefont {T.~W.}\ \bibnamefont {Hänsch}}, \bibinfo {author}
  {\bibfnamefont {A.}~\bibnamefont {Sinatra}}, \ and\ \bibinfo {author}
  {\bibfnamefont {P.}~\bibnamefont {Treutlein}},\ }\href {\doibase
  10.1038/nature08988} {\bibfield  {journal} {\bibinfo  {journal} {Nature}\
  }\textbf {\bibinfo {volume} {464}},\ \bibinfo {pages} {1170–1173} (\bibinfo
  {year} {2010})}\BibitemShut {NoStop}%
\bibitem [{\citenamefont {Gross}\ \emph {et~al.}(2010)\citenamefont {Gross},
  \citenamefont {Zibold}, \citenamefont {Nicklas}, \citenamefont {Estève},\
  and\ \citenamefont {Oberthaler}}]{Oberthaler2010}%
  \BibitemOpen
  \bibfield  {author} {\bibinfo {author} {\bibfnamefont {C.}~\bibnamefont
  {Gross}}, \bibinfo {author} {\bibfnamefont {T.}~\bibnamefont {Zibold}},
  \bibinfo {author} {\bibfnamefont {E.}~\bibnamefont {Nicklas}}, \bibinfo
  {author} {\bibfnamefont {J.}~\bibnamefont {Estève}}, \ and\ \bibinfo
  {author} {\bibfnamefont {M.~K.}\ \bibnamefont {Oberthaler}},\ }\href
  {\doibase 10.1038/nature08919} {\bibfield  {journal} {\bibinfo  {journal}
  {Nature}\ }\textbf {\bibinfo {volume} {464}},\ \bibinfo {pages} {1165–1169}
  (\bibinfo {year} {2010})}\BibitemShut {NoStop}%
\bibitem [{\citenamefont {Hamley}\ \emph {et~al.}(2012)\citenamefont {Hamley},
  \citenamefont {Gerving}, \citenamefont {Hoang}, \citenamefont {Bookjans},\
  and\ \citenamefont {Chapman}}]{Chapman2012}%
  \BibitemOpen
  \bibfield  {author} {\bibinfo {author} {\bibfnamefont {C.~D.}\ \bibnamefont
  {Hamley}}, \bibinfo {author} {\bibfnamefont {C.~S.}\ \bibnamefont {Gerving}},
  \bibinfo {author} {\bibfnamefont {T.~M.}\ \bibnamefont {Hoang}}, \bibinfo
  {author} {\bibfnamefont {E.~M.}\ \bibnamefont {Bookjans}}, \ and\ \bibinfo
  {author} {\bibfnamefont {M.~S.}\ \bibnamefont {Chapman}},\ }\href {\doibase
  10.1038/nphys2245} {\bibfield  {journal} {\bibinfo  {journal} {Nature
  Physics}\ }\textbf {\bibinfo {volume} {8}},\ \bibinfo {pages} {305–308}
  (\bibinfo {year} {2012})}\BibitemShut {NoStop}%
\bibitem [{\citenamefont {Qu}\ \emph {et~al.}(2020)\citenamefont {Qu},
  \citenamefont {Evrard}, \citenamefont {Dalibard},\ and\ \citenamefont
  {Gerbier}}]{PhysRevLett.125.033401}%
  \BibitemOpen
  \bibfield  {author} {\bibinfo {author} {\bibfnamefont {A.}~\bibnamefont
  {Qu}}, \bibinfo {author} {\bibfnamefont {B.}~\bibnamefont {Evrard}}, \bibinfo
  {author} {\bibfnamefont {J.}~\bibnamefont {Dalibard}}, \ and\ \bibinfo
  {author} {\bibfnamefont {F.}~\bibnamefont {Gerbier}},\ }\href {\doibase
  10.1103/PhysRevLett.125.033401} {\bibfield  {journal} {\bibinfo  {journal}
  {Phys. Rev. Lett.}\ }\textbf {\bibinfo {volume} {125}},\ \bibinfo {pages}
  {033401} (\bibinfo {year} {2020})}\BibitemShut {NoStop}%
\bibitem [{\citenamefont {Leroux}\ \emph {et~al.}(2010)\citenamefont {Leroux},
  \citenamefont {Schleier-Smith},\ and\ \citenamefont
  {Vuleti\ifmmode~\acute{c}\else \'{c}\fi{}}}]{PhysRevLett.104.073602}%
  \BibitemOpen
  \bibfield  {author} {\bibinfo {author} {\bibfnamefont {I.~D.}\ \bibnamefont
  {Leroux}}, \bibinfo {author} {\bibfnamefont {M.~H.}\ \bibnamefont
  {Schleier-Smith}}, \ and\ \bibinfo {author} {\bibfnamefont {V.}~\bibnamefont
  {Vuleti\ifmmode~\acute{c}\else \'{c}\fi{}}},\ }\href
  {https://link.aps.org/doi/10.1103/PhysRevLett.104.073602} {\bibfield
  {journal} {\bibinfo  {journal} {Phys. Rev. Lett.}\ }\textbf {\bibinfo
  {volume} {104}},\ \bibinfo {pages} {073602} (\bibinfo {year}
  {2010})}\BibitemShut {NoStop}%
\bibitem [{\citenamefont {Maussang}\ \emph {et~al.}(2010)\citenamefont
  {Maussang}, \citenamefont {Marti}, \citenamefont {Schneider}, \citenamefont
  {Treutlein}, \citenamefont {Li}, \citenamefont {Sinatra}, \citenamefont
  {Long}, \citenamefont {Est\`eve},\ and\ \citenamefont
  {Reichel}}]{PhysRevLett.105.080403}%
  \BibitemOpen
  \bibfield  {author} {\bibinfo {author} {\bibfnamefont {K.}~\bibnamefont
  {Maussang}}, \bibinfo {author} {\bibfnamefont {G.~E.}\ \bibnamefont {Marti}},
  \bibinfo {author} {\bibfnamefont {T.}~\bibnamefont {Schneider}}, \bibinfo
  {author} {\bibfnamefont {P.}~\bibnamefont {Treutlein}}, \bibinfo {author}
  {\bibfnamefont {Y.}~\bibnamefont {Li}}, \bibinfo {author} {\bibfnamefont
  {A.}~\bibnamefont {Sinatra}}, \bibinfo {author} {\bibfnamefont
  {R.}~\bibnamefont {Long}}, \bibinfo {author} {\bibfnamefont {J.}~\bibnamefont
  {Est\`eve}}, \ and\ \bibinfo {author} {\bibfnamefont {J.}~\bibnamefont
  {Reichel}},\ }\href {\doibase 10.1103/PhysRevLett.105.080403} {\bibfield
  {journal} {\bibinfo  {journal} {Phys. Rev. Lett.}\ }\textbf {\bibinfo
  {volume} {105}},\ \bibinfo {pages} {080403} (\bibinfo {year}
  {2010})}\BibitemShut {NoStop}%
\bibitem [{\citenamefont {Kajtoch}\ \emph {et~al.}(2018)\citenamefont
  {Kajtoch}, \citenamefont {Witkowska},\ and\ \citenamefont
  {Sinatra}}]{Kajtoch2018}%
  \BibitemOpen
  \bibfield  {author} {\bibinfo {author} {\bibfnamefont {D.}~\bibnamefont
  {Kajtoch}}, \bibinfo {author} {\bibfnamefont {E.}~\bibnamefont {Witkowska}},
  \ and\ \bibinfo {author} {\bibfnamefont {A.}~\bibnamefont {Sinatra}},\ }\href
  {\doibase 10.1209/0295-5075/123/20012} {\bibfield  {journal} {\bibinfo
  {journal} {EPL (Europhysics Letters)}\ }\textbf {\bibinfo {volume} {123}},\
  \bibinfo {pages} {20012} (\bibinfo {year} {2018})}\BibitemShut {NoStop}%
\bibitem [{\citenamefont {He}\ \emph {et~al.}(2019{\natexlab{a}})\citenamefont
  {He}, \citenamefont {Perlin}, \citenamefont {Muleady}, \citenamefont
  {Lewis-Swan}, \citenamefont {Hutson}, \citenamefont {Ye},\ and\ \citenamefont
  {Rey}}]{PhysRevResearch.1.033075}%
  \BibitemOpen
  \bibfield  {author} {\bibinfo {author} {\bibfnamefont {P.}~\bibnamefont
  {He}}, \bibinfo {author} {\bibfnamefont {M.~A.}\ \bibnamefont {Perlin}},
  \bibinfo {author} {\bibfnamefont {S.~R.}\ \bibnamefont {Muleady}}, \bibinfo
  {author} {\bibfnamefont {R.~J.}\ \bibnamefont {Lewis-Swan}}, \bibinfo
  {author} {\bibfnamefont {R.~B.}\ \bibnamefont {Hutson}}, \bibinfo {author}
  {\bibfnamefont {J.}~\bibnamefont {Ye}}, \ and\ \bibinfo {author}
  {\bibfnamefont {A.~M.}\ \bibnamefont {Rey}},\ }\href {\doibase
  10.1103/PhysRevResearch.1.033075} {\bibfield  {journal} {\bibinfo  {journal}
  {Phys. Rev. Research}\ }\textbf {\bibinfo {volume} {1}},\ \bibinfo {pages}
  {033075} (\bibinfo {year} {2019}{\natexlab{a}})}\BibitemShut {NoStop}%
\bibitem [{\citenamefont {P\l{}odzie\ifmmode~\acute{n}\else \'{n}\fi{}}\ \emph
  {et~al.}(2020)\citenamefont {P\l{}odzie\ifmmode~\acute{n}\else \'{n}\fi{}},
  \citenamefont {Ko\ifmmode~\acute{s}\else \'{s}\fi{}cielski}, \citenamefont
  {Witkowska},\ and\ \citenamefont {Sinatra}}]{Plodzien2020}%
  \BibitemOpen
  \bibfield  {author} {\bibinfo {author} {\bibfnamefont {M.}~\bibnamefont
  {P\l{}odzie\ifmmode~\acute{n}\else \'{n}\fi{}}}, \bibinfo {author}
  {\bibfnamefont {M.}~\bibnamefont {Ko\ifmmode~\acute{s}\else
  \'{s}\fi{}cielski}}, \bibinfo {author} {\bibfnamefont {E.}~\bibnamefont
  {Witkowska}}, \ and\ \bibinfo {author} {\bibfnamefont {A.}~\bibnamefont
  {Sinatra}},\ }\href {\doibase 10.1103/PhysRevA.102.013328} {\bibfield
  {journal} {\bibinfo  {journal} {Phys. Rev. A}\ }\textbf {\bibinfo {volume}
  {102}},\ \bibinfo {pages} {013328} (\bibinfo {year} {2020})}\BibitemShut
  {NoStop}%
\bibitem [{\citenamefont {Mamaev}\ \emph
  {et~al.}(2021{\natexlab{a}})\citenamefont {Mamaev}, \citenamefont {Kimchi},
  \citenamefont {Nandkishore},\ and\ \citenamefont
  {Rey}}]{PhysRevResearch.3.013178}%
  \BibitemOpen
  \bibfield  {author} {\bibinfo {author} {\bibfnamefont {M.}~\bibnamefont
  {Mamaev}}, \bibinfo {author} {\bibfnamefont {I.}~\bibnamefont {Kimchi}},
  \bibinfo {author} {\bibfnamefont {R.~M.}\ \bibnamefont {Nandkishore}}, \ and\
  \bibinfo {author} {\bibfnamefont {A.~M.}\ \bibnamefont {Rey}},\ }\href
  {\doibase 10.1103/PhysRevResearch.3.013178} {\bibfield  {journal} {\bibinfo
  {journal} {Phys. Rev. Research}\ }\textbf {\bibinfo {volume} {3}},\ \bibinfo
  {pages} {013178} (\bibinfo {year} {2021}{\natexlab{a}})}\BibitemShut
  {NoStop}%
\bibitem [{\citenamefont {Wall}\ \emph {et~al.}(2016)\citenamefont {Wall},
  \citenamefont {Koller}, \citenamefont {Li}, \citenamefont {Zhang},
  \citenamefont {Cooper}, \citenamefont {Ye},\ and\ \citenamefont
  {Rey}}]{Wall2016}%
  \BibitemOpen
  \bibfield  {author} {\bibinfo {author} {\bibfnamefont {M.~L.}\ \bibnamefont
  {Wall}}, \bibinfo {author} {\bibfnamefont {A.~P.}\ \bibnamefont {Koller}},
  \bibinfo {author} {\bibfnamefont {S.}~\bibnamefont {Li}}, \bibinfo {author}
  {\bibfnamefont {X.}~\bibnamefont {Zhang}}, \bibinfo {author} {\bibfnamefont
  {N.~R.}\ \bibnamefont {Cooper}}, \bibinfo {author} {\bibfnamefont
  {J.}~\bibnamefont {Ye}}, \ and\ \bibinfo {author} {\bibfnamefont {A.~M.}\
  \bibnamefont {Rey}},\ }\href {\doibase 10.1103/PhysRevLett.116.035301}
  {\bibfield  {journal} {\bibinfo  {journal} {Phys. Rev. Lett.}\ }\textbf
  {\bibinfo {volume} {116}},\ \bibinfo {pages} {035301} (\bibinfo {year}
  {2016})}\BibitemShut {NoStop}%
\bibitem [{\citenamefont {Kolkowitz}\ \emph
  {et~al.}(2017{\natexlab{a}})\citenamefont {Kolkowitz}, \citenamefont
  {Bromley}, \citenamefont {Bothwell}, \citenamefont {Wall}, \citenamefont
  {Marti}, \citenamefont {Koller}, \citenamefont {Zhang}, \citenamefont {Rey},\
  and\ \citenamefont {Ye}}]{Ye2017}%
  \BibitemOpen
  \bibfield  {author} {\bibinfo {author} {\bibfnamefont {S.}~\bibnamefont
  {Kolkowitz}}, \bibinfo {author} {\bibfnamefont {S.~L.}\ \bibnamefont
  {Bromley}}, \bibinfo {author} {\bibfnamefont {T.}~\bibnamefont {Bothwell}},
  \bibinfo {author} {\bibfnamefont {M.~L.}\ \bibnamefont {Wall}}, \bibinfo
  {author} {\bibfnamefont {G.~E.}\ \bibnamefont {Marti}}, \bibinfo {author}
  {\bibfnamefont {A.~P.}\ \bibnamefont {Koller}}, \bibinfo {author}
  {\bibfnamefont {X.}~\bibnamefont {Zhang}}, \bibinfo {author} {\bibfnamefont
  {A.~M.}\ \bibnamefont {Rey}}, \ and\ \bibinfo {author} {\bibfnamefont
  {J.}~\bibnamefont {Ye}},\ }\href {\doibase 10.1038/nature20811} {\bibfield
  {journal} {\bibinfo  {journal} {Nature}\ }\textbf {\bibinfo {volume} {542}},\
  \bibinfo {pages} {66} (\bibinfo {year} {2017}{\natexlab{a}})}\BibitemShut
  {NoStop}%
\bibitem [{\citenamefont {Kolkowitz}\ \emph
  {et~al.}(2017{\natexlab{b}})\citenamefont {Kolkowitz}, \citenamefont
  {Bromley}, \citenamefont {Bothwell}, \citenamefont {Wall}, \citenamefont
  {Marti}, \citenamefont {Koller}, \citenamefont {Zhang}, \citenamefont {Rey},\
  and\ \citenamefont {Ye}}]{Kolkowitz2017}%
  \BibitemOpen
  \bibfield  {author} {\bibinfo {author} {\bibfnamefont {S.}~\bibnamefont
  {Kolkowitz}}, \bibinfo {author} {\bibfnamefont {S.~L.}\ \bibnamefont
  {Bromley}}, \bibinfo {author} {\bibfnamefont {T.}~\bibnamefont {Bothwell}},
  \bibinfo {author} {\bibfnamefont {M.~L.}\ \bibnamefont {Wall}}, \bibinfo
  {author} {\bibfnamefont {G.~E.}\ \bibnamefont {Marti}}, \bibinfo {author}
  {\bibfnamefont {A.~P.}\ \bibnamefont {Koller}}, \bibinfo {author}
  {\bibfnamefont {X.}~\bibnamefont {Zhang}}, \bibinfo {author} {\bibfnamefont
  {A.~M.}\ \bibnamefont {Rey}}, \ and\ \bibinfo {author} {\bibfnamefont
  {J.}~\bibnamefont {Ye}},\ }\href {\doibase 10.1038/nature20811} {\bibfield
  {journal} {\bibinfo  {journal} {Nature}\ }\textbf {\bibinfo {volume} {542}},\
  \bibinfo {pages} {66} (\bibinfo {year} {2017}{\natexlab{b}})}\BibitemShut
  {NoStop}%
\bibitem [{\citenamefont {Bromley}\ \emph {et~al.}(2018)\citenamefont
  {Bromley}, \citenamefont {Kolkowitz}, \citenamefont {Bothwell}, \citenamefont
  {Kedar}, \citenamefont {Safavi-Naini}, \citenamefont {Wall}, \citenamefont
  {Salomon}, \citenamefont {Rey},\ and\ \citenamefont {Ye}}]{Bromley2018}%
  \BibitemOpen
  \bibfield  {author} {\bibinfo {author} {\bibfnamefont {S.~L.}\ \bibnamefont
  {Bromley}}, \bibinfo {author} {\bibfnamefont {S.}~\bibnamefont {Kolkowitz}},
  \bibinfo {author} {\bibfnamefont {T.}~\bibnamefont {Bothwell}}, \bibinfo
  {author} {\bibfnamefont {D.}~\bibnamefont {Kedar}}, \bibinfo {author}
  {\bibfnamefont {A.}~\bibnamefont {Safavi-Naini}}, \bibinfo {author}
  {\bibfnamefont {M.~L.}\ \bibnamefont {Wall}}, \bibinfo {author}
  {\bibfnamefont {C.}~\bibnamefont {Salomon}}, \bibinfo {author} {\bibfnamefont
  {A.~M.}\ \bibnamefont {Rey}}, \ and\ \bibinfo {author} {\bibfnamefont
  {J.}~\bibnamefont {Ye}},\ }\href {\doibase 10.1038/s41567-017-0029-0}
  {\bibfield  {journal} {\bibinfo  {journal} {Nature Physics}\ }\textbf
  {\bibinfo {volume} {14}},\ \bibinfo {pages} {399–404} (\bibinfo {year}
  {2018})}\BibitemShut {NoStop}%
\bibitem [{\citenamefont {He}\ \emph {et~al.}(2019{\natexlab{b}})\citenamefont
  {He}, \citenamefont {Perlin}, \citenamefont {Muleady}, \citenamefont
  {Lewis-Swan}, \citenamefont {Hutson}, \citenamefont {Ye},\ and\ \citenamefont
  {Rey}}]{He2019}%
  \BibitemOpen
  \bibfield  {author} {\bibinfo {author} {\bibfnamefont {P.}~\bibnamefont
  {He}}, \bibinfo {author} {\bibfnamefont {M.~A.}\ \bibnamefont {Perlin}},
  \bibinfo {author} {\bibfnamefont {S.~R.}\ \bibnamefont {Muleady}}, \bibinfo
  {author} {\bibfnamefont {R.~J.}\ \bibnamefont {Lewis-Swan}}, \bibinfo
  {author} {\bibfnamefont {R.~B.}\ \bibnamefont {Hutson}}, \bibinfo {author}
  {\bibfnamefont {J.}~\bibnamefont {Ye}}, \ and\ \bibinfo {author}
  {\bibfnamefont {A.~M.}\ \bibnamefont {Rey}},\ }\href {\doibase
  10.1103/PhysRevResearch.1.033075} {\bibfield  {journal} {\bibinfo  {journal}
  {Phys. Rev. Research}\ }\textbf {\bibinfo {volume} {1}},\ \bibinfo {pages}
  {033075} (\bibinfo {year} {2019}{\natexlab{b}})}\BibitemShut {NoStop}%
\bibitem [{\citenamefont {Mamaev}\ \emph
  {et~al.}(2021{\natexlab{b}})\citenamefont {Mamaev}, \citenamefont {Kimchi},
  \citenamefont {Nandkishore},\ and\ \citenamefont {Rey}}]{Mamaev2021}%
  \BibitemOpen
  \bibfield  {author} {\bibinfo {author} {\bibfnamefont {M.}~\bibnamefont
  {Mamaev}}, \bibinfo {author} {\bibfnamefont {I.}~\bibnamefont {Kimchi}},
  \bibinfo {author} {\bibfnamefont {R.~M.}\ \bibnamefont {Nandkishore}}, \ and\
  \bibinfo {author} {\bibfnamefont {A.~M.}\ \bibnamefont {Rey}},\ }\href
  {\doibase 10.1103/PhysRevResearch.3.013178} {\bibfield  {journal} {\bibinfo
  {journal} {Phys. Rev. Research}\ }\textbf {\bibinfo {volume} {3}},\ \bibinfo
  {pages} {013178} (\bibinfo {year} {2021}{\natexlab{b}})}\BibitemShut
  {NoStop}%
\bibitem [{\citenamefont {Yanes}\ \emph {et~al.}(2022)\citenamefont {Yanes},
  \citenamefont {Płodzień}, \citenamefont {Sinkevičienė}, \citenamefont
  {Žlabys}, \citenamefont {Juzeliūnas},\ and\ \citenamefont
  {Witkowska}}]{Hernandez2022}%
  \BibitemOpen
  \bibfield  {author} {\bibinfo {author} {\bibfnamefont {T.~H.}\ \bibnamefont
  {Yanes}}, \bibinfo {author} {\bibfnamefont {M.}~\bibnamefont {Płodzień}},
  \bibinfo {author} {\bibfnamefont {M.~M.}\ \bibnamefont {Sinkevičienė}},
  \bibinfo {author} {\bibfnamefont {G.}~\bibnamefont {Žlabys}}, \bibinfo
  {author} {\bibfnamefont {G.}~\bibnamefont {Juzeliūnas}}, \ and\ \bibinfo
  {author} {\bibfnamefont {E.}~\bibnamefont {Witkowska}},\ }\href {\doibase
  10.48550/ARXIV.2204.06065} {\  (\bibinfo {year} {2022}),\
  10.48550/ARXIV.2204.06065}\BibitemShut {NoStop}%
\bibitem [{\citenamefont {Civitarese}\ \emph {et~al.}(2010)\citenamefont
  {Civitarese}, \citenamefont {Reboiro}, \citenamefont {Rebon},\ and\
  \citenamefont {Tielas}}]{Civitarese2010}%
  \BibitemOpen
  \bibfield  {author} {\bibinfo {author} {\bibfnamefont {O.}~\bibnamefont
  {Civitarese}}, \bibinfo {author} {\bibfnamefont {M.}~\bibnamefont {Reboiro}},
  \bibinfo {author} {\bibfnamefont {L.}~\bibnamefont {Rebon}}, \ and\ \bibinfo
  {author} {\bibfnamefont {D.}~\bibnamefont {Tielas}},\ }\href {\doibase
  https://doi.org/10.1016/j.physleta.2009.11.013} {\bibfield  {journal}
  {\bibinfo  {journal} {Physics Letters A}\ }\textbf {\bibinfo {volume}
  {374}},\ \bibinfo {pages} {424} (\bibinfo {year} {2010})}\BibitemShut
  {NoStop}%
\bibitem [{\citenamefont {Perlin}\ \emph {et~al.}(2020)\citenamefont {Perlin},
  \citenamefont {Qu},\ and\ \citenamefont {Rey}}]{Perlin2020}%
  \BibitemOpen
  \bibfield  {author} {\bibinfo {author} {\bibfnamefont {M.~A.}\ \bibnamefont
  {Perlin}}, \bibinfo {author} {\bibfnamefont {C.}~\bibnamefont {Qu}}, \ and\
  \bibinfo {author} {\bibfnamefont {A.~M.}\ \bibnamefont {Rey}},\ }\href
  {\doibase 10.1103/PhysRevLett.125.223401} {\bibfield  {journal} {\bibinfo
  {journal} {Phys. Rev. Lett.}\ }\textbf {\bibinfo {volume} {125}},\ \bibinfo
  {pages} {223401} (\bibinfo {year} {2020})}\BibitemShut {NoStop}%
\bibitem [{\citenamefont {Bilitewski}\ \emph {et~al.}(2021)\citenamefont
  {Bilitewski}, \citenamefont {De~Marco}, \citenamefont {Li}, \citenamefont
  {Matsuda}, \citenamefont {Tobias}, \citenamefont {Valtolina}, \citenamefont
  {Ye},\ and\ \citenamefont {Rey}}]{Bilitewski2021}%
  \BibitemOpen
  \bibfield  {author} {\bibinfo {author} {\bibfnamefont {T.}~\bibnamefont
  {Bilitewski}}, \bibinfo {author} {\bibfnamefont {L.}~\bibnamefont
  {De~Marco}}, \bibinfo {author} {\bibfnamefont {J.-R.}\ \bibnamefont {Li}},
  \bibinfo {author} {\bibfnamefont {K.}~\bibnamefont {Matsuda}}, \bibinfo
  {author} {\bibfnamefont {W.~G.}\ \bibnamefont {Tobias}}, \bibinfo {author}
  {\bibfnamefont {G.}~\bibnamefont {Valtolina}}, \bibinfo {author}
  {\bibfnamefont {J.}~\bibnamefont {Ye}}, \ and\ \bibinfo {author}
  {\bibfnamefont {A.~M.}\ \bibnamefont {Rey}},\ }\href {\doibase
  10.1103/PhysRevLett.126.113401} {\bibfield  {journal} {\bibinfo  {journal}
  {Phys. Rev. Lett.}\ }\textbf {\bibinfo {volume} {126}},\ \bibinfo {pages}
  {113401} (\bibinfo {year} {2021})}\BibitemShut {NoStop}%
\bibitem [{\citenamefont {Roscilde}\ \emph {et~al.}(2021)\citenamefont
  {Roscilde}, \citenamefont {Mezzacapo},\ and\ \citenamefont
  {Comparin}}]{Roscilde2021}%
  \BibitemOpen
  \bibfield  {author} {\bibinfo {author} {\bibfnamefont {T.}~\bibnamefont
  {Roscilde}}, \bibinfo {author} {\bibfnamefont {F.}~\bibnamefont {Mezzacapo}},
  \ and\ \bibinfo {author} {\bibfnamefont {T.}~\bibnamefont {Comparin}},\
  }\href {\doibase 10.1103/PhysRevA.104.L040601} {\bibfield  {journal}
  {\bibinfo  {journal} {Phys. Rev. A}\ }\textbf {\bibinfo {volume} {104}},\
  \bibinfo {pages} {L040601} (\bibinfo {year} {2021})}\BibitemShut {NoStop}%
\bibitem [{\citenamefont {Wu}\ \emph {et~al.}(2022)\citenamefont {Wu},
  \citenamefont {Lin}, \citenamefont {Ding}, \citenamefont {Zheng},
  \citenamefont {Lesanovsky},\ and\ \citenamefont {Li}}]{Wu2022}%
  \BibitemOpen
  \bibfield  {author} {\bibinfo {author} {\bibfnamefont {H.}~\bibnamefont
  {Wu}}, \bibinfo {author} {\bibfnamefont {X.-Y.}\ \bibnamefont {Lin}},
  \bibinfo {author} {\bibfnamefont {Z.-X.}\ \bibnamefont {Ding}}, \bibinfo
  {author} {\bibfnamefont {S.-B.}\ \bibnamefont {Zheng}}, \bibinfo {author}
  {\bibfnamefont {I.}~\bibnamefont {Lesanovsky}}, \ and\ \bibinfo {author}
  {\bibfnamefont {W.}~\bibnamefont {Li}},\ }\href@noop {} {\bibfield  {journal}
  {\bibinfo  {journal} {Science China Physics, Mechanics {\&} Astronomy}\
  }\textbf {\bibinfo {volume} {65}},\ \bibinfo {pages} {280311} (\bibinfo
  {year} {2022})}\BibitemShut {NoStop}%
\bibitem [{\citenamefont {Comparin}\ \emph
  {et~al.}(2022{\natexlab{a}})\citenamefont {Comparin}, \citenamefont
  {Mezzacapo},\ and\ \citenamefont {Roscilde}}]{Comparin2022_a}%
  \BibitemOpen
  \bibfield  {author} {\bibinfo {author} {\bibfnamefont {T.}~\bibnamefont
  {Comparin}}, \bibinfo {author} {\bibfnamefont {F.}~\bibnamefont {Mezzacapo}},
  \ and\ \bibinfo {author} {\bibfnamefont {T.}~\bibnamefont {Roscilde}},\
  }\href {\doibase 10.48550/ARXIV.2205.03910} {\enquote {\bibinfo {title}
  {Multipartite entangled states in dipolar quantum simulators},}\ } (\bibinfo
  {year} {2022}{\natexlab{a}})\BibitemShut {NoStop}%
\bibitem [{\citenamefont {Comparin}\ \emph
  {et~al.}(2022{\natexlab{b}})\citenamefont {Comparin}, \citenamefont
  {Mezzacapo},\ and\ \citenamefont {Roscilde}}]{Comparin2022_b}%
  \BibitemOpen
  \bibfield  {author} {\bibinfo {author} {\bibfnamefont {T.}~\bibnamefont
  {Comparin}}, \bibinfo {author} {\bibfnamefont {F.}~\bibnamefont {Mezzacapo}},
  \ and\ \bibinfo {author} {\bibfnamefont {T.}~\bibnamefont {Roscilde}},\
  }\href {\doibase 10.1103/PhysRevA.105.022625} {\bibfield  {journal} {\bibinfo
   {journal} {Phys. Rev. A}\ }\textbf {\bibinfo {volume} {105}},\ \bibinfo
  {pages} {022625} (\bibinfo {year} {2022}{\natexlab{b}})}\BibitemShut
  {NoStop}%
\bibitem [{\citenamefont {Anglin}\ and\ \citenamefont
  {Vardi}(2001)}]{PhysRevA.64.013605}%
  \BibitemOpen
  \bibfield  {author} {\bibinfo {author} {\bibfnamefont {J.~R.}\ \bibnamefont
  {Anglin}}\ and\ \bibinfo {author} {\bibfnamefont {A.}~\bibnamefont {Vardi}},\
  }\href {\doibase 10.1103/PhysRevA.64.013605} {\bibfield  {journal} {\bibinfo
  {journal} {Phys. Rev. A}\ }\textbf {\bibinfo {volume} {64}},\ \bibinfo
  {pages} {013605} (\bibinfo {year} {2001})}\BibitemShut {NoStop}%
\bibitem [{\citenamefont {Andr\'e}\ and\ \citenamefont
  {Lukin}(2002)}]{PhysRevA.65.053819}%
  \BibitemOpen
  \bibfield  {author} {\bibinfo {author} {\bibfnamefont {A.}~\bibnamefont
  {Andr\'e}}\ and\ \bibinfo {author} {\bibfnamefont {M.~D.}\ \bibnamefont
  {Lukin}},\ }\href {\doibase 10.1103/PhysRevA.65.053819} {\bibfield  {journal}
  {\bibinfo  {journal} {Phys. Rev. A}\ }\textbf {\bibinfo {volume} {65}},\
  \bibinfo {pages} {053819} (\bibinfo {year} {2002})}\BibitemShut {NoStop}%
\bibitem [{\citenamefont {Dutta}\ \emph {et~al.}(2015)\citenamefont {Dutta},
  \citenamefont {Gajda}, \citenamefont {Hauke}, \citenamefont {Lewenstein},
  \citenamefont {Lühmann}, \citenamefont {Malomed}, \citenamefont
  {Sowi{\'{n}}ski},\ and\ \citenamefont {Zakrzewski}}]{nonstandard}%
  \BibitemOpen
  \bibfield  {author} {\bibinfo {author} {\bibfnamefont {O.}~\bibnamefont
  {Dutta}}, \bibinfo {author} {\bibfnamefont {M.}~\bibnamefont {Gajda}},
  \bibinfo {author} {\bibfnamefont {P.}~\bibnamefont {Hauke}}, \bibinfo
  {author} {\bibfnamefont {M.}~\bibnamefont {Lewenstein}}, \bibinfo {author}
  {\bibfnamefont {D.-S.}\ \bibnamefont {Lühmann}}, \bibinfo {author}
  {\bibfnamefont {B.~A.}\ \bibnamefont {Malomed}}, \bibinfo {author}
  {\bibfnamefont {T.}~\bibnamefont {Sowi{\'{n}}ski}}, \ and\ \bibinfo {author}
  {\bibfnamefont {J.}~\bibnamefont {Zakrzewski}},\ }\href {\doibase
  10.1088/0034-4885/78/6/066001} {\bibfield  {journal} {\bibinfo  {journal}
  {Reports on Progress in Physics}\ }\textbf {\bibinfo {volume} {78}},\
  \bibinfo {pages} {066001} (\bibinfo {year} {2015})}\BibitemShut {NoStop}%
\bibitem [{\citenamefont {Trimborn}\ \emph {et~al.}(2009)\citenamefont
  {Trimborn}, \citenamefont {Witthaut},\ and\ \citenamefont
  {Korsch}}]{PhysRevA.79.013608}%
  \BibitemOpen
  \bibfield  {author} {\bibinfo {author} {\bibfnamefont {F.}~\bibnamefont
  {Trimborn}}, \bibinfo {author} {\bibfnamefont {D.}~\bibnamefont {Witthaut}},
  \ and\ \bibinfo {author} {\bibfnamefont {H.~J.}\ \bibnamefont {Korsch}},\
  }\href {\doibase 10.1103/PhysRevA.79.013608} {\bibfield  {journal} {\bibinfo
  {journal} {Phys. Rev. A}\ }\textbf {\bibinfo {volume} {79}},\ \bibinfo
  {pages} {013608} (\bibinfo {year} {2009})}\BibitemShut {NoStop}%
\bibitem [{\citenamefont {Shchesnovich}\ and\ \citenamefont
  {Konotop}(2008)}]{PhysRevA.77.013614}%
  \BibitemOpen
  \bibfield  {author} {\bibinfo {author} {\bibfnamefont {V.~S.}\ \bibnamefont
  {Shchesnovich}}\ and\ \bibinfo {author} {\bibfnamefont {V.~V.}\ \bibnamefont
  {Konotop}},\ }\href {\doibase 10.1103/PhysRevA.77.013614} {\bibfield
  {journal} {\bibinfo  {journal} {Phys. Rev. A}\ }\textbf {\bibinfo {volume}
  {77}},\ \bibinfo {pages} {013614} (\bibinfo {year} {2008})}\BibitemShut
  {NoStop}%
\bibitem [{\citenamefont {Smerzi}\ \emph {et~al.}(1997)\citenamefont {Smerzi},
  \citenamefont {Fantoni}, \citenamefont {Giovanazzi},\ and\ \citenamefont
  {Shenoy}}]{PhysRevLett.79.4950}%
  \BibitemOpen
  \bibfield  {author} {\bibinfo {author} {\bibfnamefont {A.}~\bibnamefont
  {Smerzi}}, \bibinfo {author} {\bibfnamefont {S.}~\bibnamefont {Fantoni}},
  \bibinfo {author} {\bibfnamefont {S.}~\bibnamefont {Giovanazzi}}, \ and\
  \bibinfo {author} {\bibfnamefont {S.~R.}\ \bibnamefont {Shenoy}},\ }\href
  {\doibase 10.1103/PhysRevLett.79.4950} {\bibfield  {journal} {\bibinfo
  {journal} {Phys. Rev. Lett.}\ }\textbf {\bibinfo {volume} {79}},\ \bibinfo
  {pages} {4950} (\bibinfo {year} {1997})}\BibitemShut {NoStop}%
\end{thebibliography}%

\end{document}